\documentclass[preprint,12pt]{elsarticle}

\usepackage{amssymb}
\usepackage{listings}
\usepackage{url}
\usepackage{graphicx}
\usepackage{dcolumn}
\usepackage{bm}
\usepackage[]{algorithm2e}
\usepackage{mathtools}
\DeclarePairedDelimiter{\ceil}{\lceil}{\rceil}
\DeclarePairedDelimiter{\floor}{\lfloor}{\rfloor}

\journal{Computer Physics Communications}

\begin{document}

\begin{frontmatter}



\title{WaveRange: Wavelet-based data compression for three-dimensional numerical simulations on regular grids}


\author[label1]{Dmitry Kolomenskiy\corref{cor1}}
\author[label1]{Ryo Onishi}
\author[label1]{Hitoshi Uehara}
\address[label1]{Japan Agency for Marine-Earth Science and Technology (JAMSTEC), Japan}
\cortext[cor1]{Corresponding author: dkolom@gmail.com; dkolomenskiy@jamstec.go.jp}

\begin{abstract}
    A wavelet-based method for compression of three-dimensional simulation data is presented and its software framework is described.
    It uses wavelet decomposition and subsequent range coding with quantization suitable for floating-point data.
    The effectiveness of this method is demonstrated by applying it to 
    example numerical tests, ranging from idealized configurations to realistic global-scale simulations.

\end{abstract}


\end{frontmatter}




\section*{Program summary}

\begin{itemize}

\item Program title: WaveRange v3.15.

\item Licensing provisions: GNU General Public License 3 (GPL).

\item Programming language: C and C++.

\item Supplementary material: Sample compressed dataset using FluSI HDF5 format \verb|data_sample.zip|.

\item Nature of problem: WaveRange compresses three-dimensional floating-point data produced by computational physics solvers. It reads the input data
stored in files as three-dimensional floating-point arrays, and generates smaller compressed files. Subsequently, WaveRange
reads compressed files and approximately reconstructs the original data approximately with controlled accuracy.

\item Solution method: Lossy data compression is achieved by application of a three-dimensional discrete wavelet transform, quantization and range coding. 
The quantization method is designed such as to ensure the required reconstruction error. Reconstruction is the inverse of the compression steps in the reverse order.

\item Additional comments including Restrictions and Unusual features: WaveRange can be used as a standalone application or as a library. Currently supported data formats in the application mode are `generic' Fortran/C/C++, FluSI and MSSG output and restart files.

\item References: All appropriate methodological references are contained in the section entitled References.

\end{itemize}

\section{\label{sec:intro}Introduction}

Partial differential equations often arise in physical sciences from three-dimensional (3D) continuum models, yielding boundary value problems for continuous field variables defined over 3D spatial domains. Numerical solution of these problems involves discretization and, among all available methods, many employ regular grids such that the discretized field variables can be stored as three-dimensional arrays. Regular grids prevail in the high-performance computing (HPC) for enabling fast and efficient implementation of high-order numerical methods with good parallel scalability. HPC simulations based on Cartesian grids are extremely diverse and include, to name a few, particle-laiden fluid flows \cite{Fornari_2016_pof}, solar flares \cite{Bakke_2016_aa}
and neutron transport inside the core of a nuclear reactor \cite{Baudron_2007_nse}.
In Earth science, curvilinear grids such as Yin-Yang \cite{Kageyama_2004_ggg} or Cubed Sphere \cite{Ronchi_1996_jcp} 
are used for global weather simulations \cite{Nakano_2017_gmd}, geodynamo simulations \cite{Kageyama_2004_proc}, calculation of seismograms \cite{Komatitsch_2013_book}, etc. 
Toroidal grids are used in tokamak plasma simulations \cite{Korpilo_2016_cpc}.

HPC produces large volumes of output data. A numerical simulation using several hundred or thousands of processor cores would allocate three-dimensional arrays totaling to several Gigabytes or Terabytes. If the solution evolves in time, a new three-dimensional data set is produced at every time step. This massive data flow is characteristic of big data applications \cite{Hilbert_2016}, and it is not surprising that high-performance numerical simulations using regular grids hit the limitations of the contemporary data handling technologies. 
In particular, data storage capacity is finite. To alleviate this constraint in practice, \emph{in situ} data reduction is routinely performed during the simulation and only the selected integral physical quantities or time-resolved sequences are stored. Nevertheless, it is often required to store the three-dimensional fields for purposes such as scientific visualization, restart of the simulation or additional post-processing. These large datasets quickly saturate the available disk space if stored as floating-point arrays without compression.

Lossless data compression tools, such as LZMA compression, reduce the typical floating-point binary data file size by less than 20\%. Accepting some data loss, it is a common practice to store the simulation output fields in single precision and downsample the data, e.g., save every second point value in each direction. Such reduced datasets, while being of insufficient information capacity for the simulation, are often suitable for postprocessing. Given that the differences between neighboring point values can be interpreted as wavelet coefficients, a more refined version of the mentioned approach is to apply wavelet transform to the field and encode the significant portion of the wavelet coefficients using a common data compression method such as entropy coding. 
This technique is currently widely in use for image compression, being part of the JPEG2000 standard \cite{Taubman_2002_book}. 
Its suitability for the computational fluid dynamics (CFD) data compression has been evaluated in \cite{Schmalzl_2003_cg}, alongside other image compression algorithms,
and later revisited in \cite{Woodring_2011_ieee}.
A related method has been recently implemented for multiresolution rendering and storage of geoscience models with discontinuities \cite{Peyrot_2019_jcg}.
In \cite{Sakai_2011_conf,Sakai_2013_candf,Sakai_2013_ijnmf}, wavelet compression has been used 
in the context of numerical simulation of industrial fluid flows using the building-cube method, with focus on aeroacoustics.
Overall good performance has been reported in terms of the compression ratio, accuracy and parallel performance for large datasets.
However, error control was not explicitly handed.

Of all types of output, the restart data may pose the most stringent accuracy constraint on the lossy compression,
because it is commonly expected that restart should not influence the final result of the simulation.
One can expect that the required restart data accuracy depend on the physical model.
Indeed, there exist models that
are insensitive to ample reduction in the width of the floating-point significand \cite{Hatfield_2018_mwr}.
The use of lossy data compression techniques has also been advocated
by showing that compression effects are often unimportant or disappear in post-processing analyses \cite{Baker_2016_gmd},
and substantial gain in the compression ratio can be achieved while keeping the error at acceptable level in terms of physically motivated metrics \cite{Laney_2013_ieee}.
Consequently, it appears reasonable to adjust the restart data storage to the precision justified by the level of model error.
We further investigate into this issue by considering two atmospheric dynamics simulations in the present paper.

In the context of fluid dynamics and atmospheric science, since the ability of wavelets to provide compressed representation of 
turbulent flows was recognized \cite{Farge_1992_arfm}, 
a significant body of research
focused on the development of wavelet-based adaptive numerical methods allowing
to lower the computational complexity and memory requirements of high-Reynolds number flows simulations \cite{Schneider_2010_arfm}.
Studies taking the perspective of CFD data storage remain relatively sparse. Besides the aforementioned work, 
a wavelet transform-vector quantization compression method for ocean models was proposed in \cite{Bradley_1993_proc},
the effect of lossy wavelet-based compression on barotropic turbulence simulation data has been studied in \cite{Wilson_2002_proc},
a hybrid method with supercompact multiwavelets was suggested in \cite{Kang_2003_ksme},
tradeoffs in accuracy, storage cost and execution times using different wavelet transforms were considered in \cite{Li_2015_ieee}.

It should be mentioned that wavelet bases are not the only 
that yield sparse representation of turbulent flow fields.
Decompositions such as POD \cite{Berkooz_1993_arfm,Balajewicz_2013_jfm} or DMD \cite{Schmid_2010_jfm} are also used for this purpose,
and employed in CFD output data compression methods \cite{Lorente_2010_ast,Bi_2014_ieee}.
Each method has its own advantages, but in this paper we only consider the wavelet-based approach that may be more suitable for large datasets for its lower computational complexity, compared with the POD or DMD. A comparison of dimensionality reduction using POD and wavelet coherent structure identification
can be found in \cite{Schlegel_2009_proc}.
The wavelet-based method presented in this work does not
require any time history, i.e., it can be applied to compress a single time snapshot.

Sub-band coding (SBC) \cite{Rosten_2004_gp}
and the use of more general filter banks than the discrete wavelet \cite{Duval_2000_proc}
have been considered in the context of seismic data compression.
Another, conceptually different family of methods can be described as prediction-based compression algorithms (see \cite{Najmabadi_2017_comput,Lakshminarasimhan_2011_europar,Liang_2018_ieee} and references therein) that exploit spatio-temporal patterns in the data.
A comparative discussion of different existing approaches to scientific data reduction, 
including lossy data compression, can be found in a recent review paper by Li~\textit{et~al.} \cite{Li_2018_cgf}.

The first objective of this work is to implement a data compression method suitable for files that contain three-dimensional floating-point arrays output from numerical simulation.
We mainly target applications in Earth science such as atmospheric dynamics simulation, but 
the method and the software are designed to fit broader use. 
Our approach is similar to \cite{Sakai_2013_candf} conceptually, but differs in many aspects such as error control, wavelet transform depth, etc.
Therefore, a self-contained description of the method is provided in Section~\ref{sec:method}.
The computer code is implemented in C/C++, it is open-source and accessible via \url{https://github.com/pseudospectators/WaveRange}.
It is described in Section~\ref{sec:software}.
Our second objective is to evaluate the performance of the method and to 
devise practical recommendations for users. This constitutes Section~\ref{sec:results} of the paper. 
We particularly focus on the relationship between reconstruction error and
compression ratio, as well as its effect on the accuracy of post-processing and simulation restart.
The compression and decompression performance with consideration of computational cost is examined in Section~\ref{sec:cost}.
Section~\ref{sec:conclu} contains concluding remarks.

\section{\label{sec:method}Problem definition and description of the method}

\begin{figure}
    \begin{center}
        \includegraphics[width=0.9\textwidth,clip]{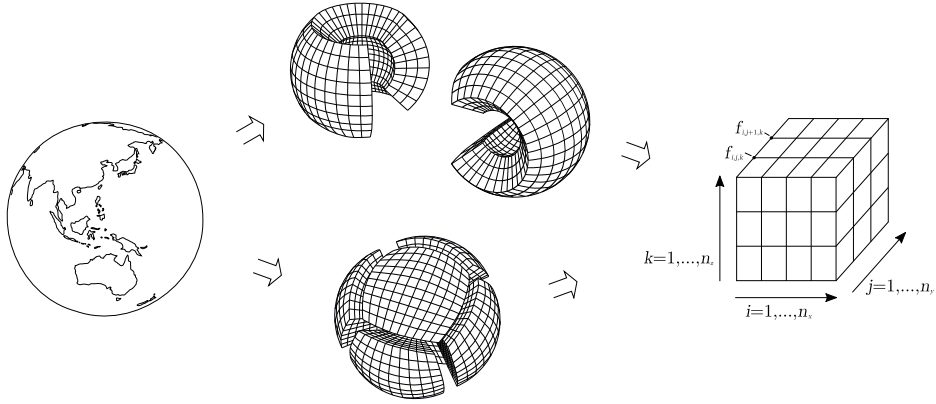}      
    \end{center}
    \caption{Schematic drawing of a Yin-Yang grid \cite{Kageyama_2004_ggg} and a Cubed Sphere grid \cite{Ronchi_1996_jcp} used in Earth science computations, 
    and a Cartesian grid block with point-value data stored at the grid nodes.}
    \label{fig:grids}
    \vspace*{-3mm}
\end{figure}

We restrict our attention to data sampled on single or multi-block grids with 
each block using three-dimensional Cartesian indexing, as shown in Fig.~\ref{fig:grids}.
Numerical methods that involve such kinds of topology are common in HPC
for the ease and efficiency of data management.
Atmospheric flow simulations
of the global scale can be performed using a Yin-Yang grid that consists of two 
overlapping blocks.
Regional and urban simulations can incorporate geometrical 
representation of landscape features and buildings by using
immersed boundary approaches \cite{Lundquist_2010_mwr,Matsuda_2018_jweia}
that effectively reduce the computational domain to a rectangular box.

Let $\{f_{i_x,i_y,i_z}\}$ be a three-dimensional scalar field sampled on a grid
$\{x_{i_x,i_y,i_z}, y_{i_x,i_y,i_z}, z_{i_x,i_y,i_z}\}$, where $i_x=1,...,n_x$; $i_y=1,...,n_y$; $i_z=1,...,n_z$. 
This grid may be curvilinear. However, the positions of the grid nodes in 
the physical space are not required by the data compression algorithm.
Only the array of values $f_{i_x,i_y,i_z}$ and the number of grid points in each direction $n_x$, $n_y$ and $n_z$
constitute the input data.
If the grid is multiblock, each block can either be treated independently by the compression algorithm, 
or the blocks can be merged in one array if their dimensions match.
Therefore, in the following discussion we will refer to
the rectilinear grid schematically shown in Fig.~\ref{fig:grids}, without loss of generality.

\begin{figure}
    \begin{center}
        \includegraphics[width=0.9\textwidth,clip]{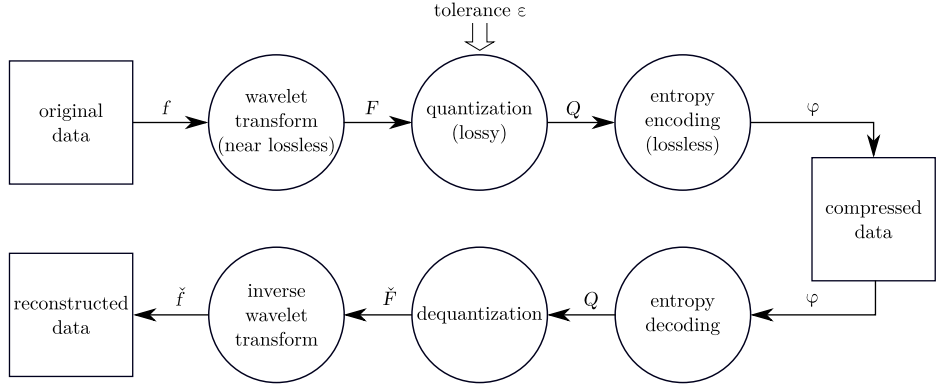}      
    \end{center}
    \caption{Data flow diagram of the compression and reconstruction method.}
    \label{fig:data_flow}
    \vspace*{-3mm}
\end{figure}

On the highest level, the compression method consists of the following steps: (I) wavelet transform;
(II) quantization;
(III) entropy coding.
Reconstruction is the inverse of the above operations in the reverse order. 
These steps are schematically shown in the data flow diagram in Fig.~\ref{fig:data_flow}.
Note that `dequantization' is not the exact inverse of the quantization, i.e., $F \ne \check{F}$ in general.

\subsection{\label{sec:wtransform}Wavelet transform}

Wavelet transform can provide compressed representation of a signal if it is locally correlated, and
it becomes particularly efficient if the fine-scale activity is sparse.
For example, turbulent flow fields satisfy these criteria, and it is known that only few wavelet coefficients are 
sufficient to represent the dynamically active part of the flow \cite{Schneider_2010_arfm}.

The wavelet transform produces an array of real values $\{F_{i_x,i_y,i_z}\}$ with the same number of elements as in 
the original array $\{f_{i_x,i_y,i_z}\}$, i.e., containing $n_x \times n_y \times n_z$ elements in total. We use a three-dimensional multiresolution transform based on the bi-orthogonal Cohen--Daubechies--Feauveau 9/7 (CDF9/7) wavelet, expressed in terms of lifting steps \cite{Daubechies_1998_jfaa,Getreuer_website}. 
Numerical experiments with sample CFD velocity data confirmed that this transform yields higher compression ratio than
using lower order biorthogonal wavelets or orthogonal wavelets such as Haar, 
Daubechies (D4...D20) and Symlets (Sym4...Sym10).
Previous work \cite{Li_2015_ieee} also showed that the CDF9/7 wavelet performed better than the Haar wavelet
for scientific data compression, albeit using thresholding of the wavelet coefficients rather than entropy coding. 

Basically, the transform consists of a finite sequence of filtering steps, 
called lifting steps,
applied to one-dimensional (1D) signal. 
Let ${g_i}$, $i=1,...,m$ be a 1D array of real values.
One level of the forward transform consists in calculating the
values of approximation coefficients ${s_j}$,
$j=1,...,\ceil{m/2}$ and detail coefficients ${d_j}$, $j=1,...,\floor{m/2}$ 
($\ceil{\cdot}$ and $\floor{\cdot}$ are the ceiling and the floor functions, respectively) using lifting steps
\begin{equation}
\begin{split}
s_j^{(0)} & = g_{2j-1}, \\
d_j^{(0)} & = g_{2j}, \\
d_j^{(1)} & = d_j^{(0)} + \alpha \left( s_j^{(0)} + s_{j+1}^{(0)} \right), \\
s_j^{(1)} & = s_j^{(0)} + \beta \left( d_j^{(1)} + d_{j-1}^{(1)} \right), \\
d_j^{(2)} & = d_j^{(1)} + \gamma \left( s_j^{(1)} + s_{j+1}^{(1)} \right), \\
s_j^{(2)} & = s_j^{(1)} + \delta \left( d_j^{(2)} + d_{j-1}^{(2)} \right), \\
s_j & = \zeta s_j^{(2)}, \\
d_j & = d_j^{(2)} / \zeta,
\end{split}
\label{eq:cdf97_lifting}
\end{equation}
where $\alpha=-1.5861343420693648$, $\beta=-0.0529801185718856$, $\gamma=0.8829110755411875$, $\delta=0.4435068520511142$ and $\zeta=1.1496043988602418$.
For boundary handling, coefficients $s_{\ceil{m/2}+1}^{(0)}$, $s_{\ceil{m/2}+1}^{(1)}$, 
$d_{0}^{(1)}$ and $d_{0}^{(2)}$ are defined and set identical to zero.
If $m$ is odd, then the 
missing element $d_{\ceil{m/2}}^{(0)}$, which is necessary for calculating $d_{\ceil{m/2}}^{(1)}$ and $s_{\ceil{m/2}}^{(1)}$ subsequently,
is determined using extrapolation
\begin{equation}
d_{\ceil{m/2}}^{(0)} = - \frac{2}{1+2\beta\gamma} \left( \alpha\beta\gamma s_{\ceil{m/2}-1}^{(0)} + \beta\gamma d_{\ceil{m/2}-1}^{(0)} + (\alpha+\gamma+3\alpha\beta\gamma) s_{\ceil{m/2}}^{(0)} \right).
\label{eq:extrp_lifting}
\end{equation}
The output of (\ref{eq:cdf97_lifting}) is an array of size $m$ in which the first $\ceil{m/2}$ elements
contain the approximation coefficients $s_j$ and the last $\floor{m/2}$ elements contain the detail coefficients $d_j$.


The inverse transform admits coefficients $s_j$ and $d_j$ at input, and resolves the lifting steps 
($\ref{eq:cdf97_lifting}$) in the reverse order to produce $g_j$ at the output.
More specifically, $s_l^{(2)}$ and $d_l^{(2)}$ are determined
from the last two lines, then the third to last equation is solved with respect to $s_l^{(1)}$ and so on.

The three-dimensional transform is constructed by applying the above one-dimensional transform
sequentially in the three directions of the three-dimensional data array.
WaveRange does one level of the 1D wavelet transform in the $x$ direction, then in $y$, and finally in $z$.
After that, it moves to the next level:
the approximation coefficients on the first level are taken as input data for 1D transforms.
Then, the same procedure is repeated on the second level, etc., as explained in Algorithm~\ref{algo:wt3d}.
We found that repeating the process 4 times is practically sufficient to reach the maximum compression ratio,
because, after 4 levels of the transform, the number of approximation coefficients becomes as small as $1/4096$
of the total number of elements in the dataset.
We therefore always set $L=4$ in WaveRange.
This means that, as long as $n_x \ge 15$, $n_y \ge 15$ and $n_z \ge 15$, 
the same number (four) of 1D transform levels is realized in all three directions,
even in those cases when there is a dramatic difference between $n_x$, $n_y$ and $n_z$.
Note that the transform is in-place, i.e., $\{f_{i_x,i_y,i_z}\}$ and $\{F_{i_x,i_y,i_z}\}$ 
physically share the same memory by being stored in the same array.

\begin{algorithm}[p]
 \SetAlgoNoLine
 \SetInd{0.5em}{0.5em}
 \KwData{3D array of point values in physical space $f_{i_x,i_y,i_z}$, $i_x=1,...,n_x$, $i_y=1,...,n_y$, $i_z=1,...,n_z$.}
 \KwResult{approximation coefficients and detail coefficients placed in the same array $F_{i_x,i_y,i_z}$.}
 $m_x \gets n_x$;
 $m_y \gets n_y$;
 $m_z \gets n_z$\;
 $\{F_{i_x,i_y,i_z}\} \equiv \{f_{i_x,i_y,i_z}\}$\;  
 \For{$l=1$ \rm{to} $L$}{
   \If{$m_x>1$}{ 
     \For{$i_y=1$ \rm{to} $m_y$\rm{;} $i_z=1$ \rm{to} $m_z$}{
       copy data elements $F_{j_x,i_y,i_z}$ in a contiguous array $g_{j_x}$, where $j_x=1,...,m_x$\;
       apply the 1D wavelet transform (\ref{eq:cdf97_lifting}) and copy the result in $F_{j_x,i_y,i_z}$\;
     }
   }
   \If{$m_y>1$}{ 
     \For{$i_x=1$ \rm{to} $m_x$\rm{;} $i_z=1$ \rm{to} $m_z$}{
       copy data elements $F_{i_x,j_y,i_z}$ in a contiguous array $g_{j_y}$, where $j_y=1,...,m_y$\;
       apply the 1D wavelet transform (\ref{eq:cdf97_lifting}) and copy the result in $F_{i_x,j_y,i_z}$\;
     }
   }
   \If{$m_z>1$}{ 
     \For{$i_x=1$ \rm{to} $m_x$\rm{;} $i_y=1$ \rm{to} $m_y$}{
       copy data elements $F_{i_x,i_y,j_z}$ in a contiguous array $g_{j_z}$, where $j_z=1,...,m_z$\;
       apply the 1D wavelet transform (\ref{eq:cdf97_lifting}) and copy the result in $F_{i_x,i_y,j_z}$\;
     }
   }
 $m_x \gets \ceil{m_x/2}$;
 $m_y \gets \ceil{m_y/2}$;
 $m_z \gets \ceil{m_z/2}$\;     
 }      
 \caption{Three-dimensional wavelet transform.}\label{algo:wt3d}
  
\end{algorithm}

The inverse transform has similar algorithmic structure.
It starts with the approximation coefficients and the details at the largest scale,
and repeats $L$ iterations adding one extra level of details in each direction on each iteration.

\begin{figure}
    \begin{center}
        \includegraphics[width=0.9\textwidth,clip]{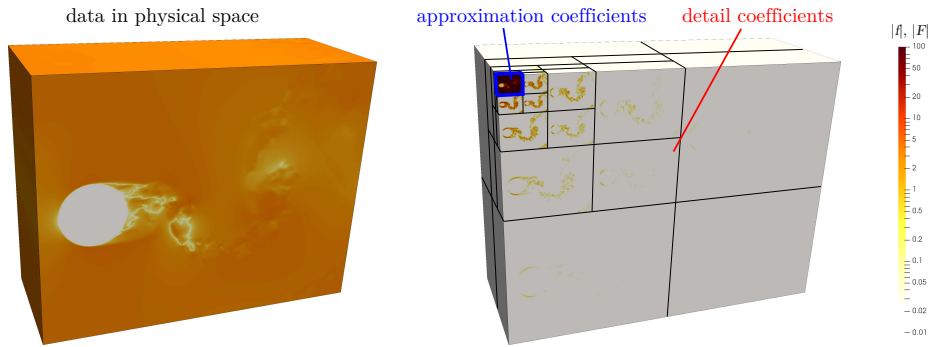}      
    \end{center}
    \caption{Example visualization of a 3D dataset and its wavelet transform.}
    \label{fig:wt3d}
    \vspace*{-3mm}
\end{figure}

An example field and its transform are displayed 
in figure~\ref{fig:wt3d}. On the left, the original 
field in physical space, $\{f_{i_x,i_y,i_z}\}$, is displayed.
In this example, it contains point values of the velocity component $u_x$
in a turbulent wake flow, which is described in Section~\ref{sec:wake_compression}.
The magnitude of the point data values is shown on a logarithmic color scale.
On the right, the output of Algorithm~\ref{algo:wt3d} is shown, 
in which the approximation coefficients and the detail coefficients on all levels
are packed in one three-dimensional array $\{F_{i_x,i_y,i_z}\}$.
Seven-eighth of its elements correspond to the smallest-scale detail coefficients.
They are all small in magnitude: most of them are below the visibility threshold of the selected color scale,
and only few large ones appear in the boundary layer.
On the next and subsequent levels, details in the turbulent wake become increasingly larger in magnitude.
The approximation coefficients occupy the upper-left corner of the domain.
They are small in number and large in magnitude.
Note that the transform is near lossless (in the terminology of \cite{Li_2018_cgf}) such that all values of $\{f_{i_x,i_y,i_z}\}$
can be calculated from $\{F_{i_x,i_y,i_z}\}$ with floating point round-off accuracy.

\subsection{\label{sec:quantization}Quantization}

From this point on, $\{F_{i_x,i_y,i_z}\}$ is treated as a one-dimensional array of real values $F_i$, $i=1,...,N$, where $N=n_x \times n_y \times n_z$.
Quantization represents each element $F_i$ as a set of 1-byte numbers (i.e., integer numbers ranging from 0 to 255), as required for the subsequent entropy coding. 
In general, entropy coders are not limited to $256$ size alphabet, but 
this size is convenient as it corresponds to the \textit{char} type, native in C language.
A double-precision floating point variable $F_i$ can be losslessly quantized using eight 1-byte numbers. 
However, in applications related to the numerical simulation of turbulent flows, quantization with some data loss has to be accepted
in order to achieve the desired high compression ratio.

\begin{figure}
    \begin{center}
        \includegraphics[width=0.8\textwidth,clip]{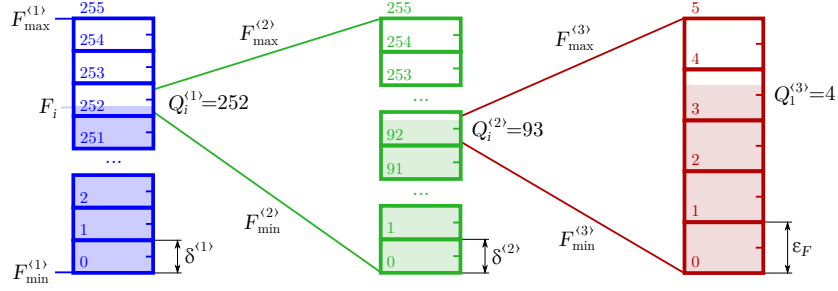}      
    \end{center}
    \caption{Illustrative example of the data quantization procedure.}
    \label{fig:quant}
    \vspace*{-3mm}
\end{figure}

Lossy compression requires less than eight 1-byte integer numbers $Q_{i}^{\langle j \rangle}$ per variable $F_i$,
indexed with a superscript $\langle j \rangle$, using the following approximation:
\begin{equation}
F_i \approx \sum_{j=1}^{J} ( Q_i^{\langle j \rangle} \delta^{\langle j \rangle} + F^{\langle j \rangle}_{\min} ), 
\label{eq:quant}
\end{equation}
with the approximation error no greater than $\delta^{\langle J \rangle}$.
The latter is controlled by the
relative tolerance $\varepsilon$, which is a user-specified parameter of the compression routine.
Note that $\varepsilon$ can be regarded as the desired $L^\infty$ error in $f$ in physical space,
normalized with $\max{|f|}$, but $\delta^{\langle J \rangle}$ in (\ref{eq:quant}) sets the absolute error in $F_i$ in wavelet space.
To relate $\delta^{\langle J \rangle}$ with $\varepsilon$, we define 
\begin{equation}
\varepsilon_F =  \varepsilon \max |f_{i_x,i_y,i_z}| / \eta,
\label{eq:tolabs}
\end{equation}
where the maximum is taken over all elements, and $\eta$ is a constant coefficient.
By trial and error we have found that the value $\eta = 1.75$ guarantees that $||f-\check{f}||_\infty / \max{|f|} \approx \varepsilon$
as long as we use four levels of the wavelet transform.
Quantization adds random noise with amplitude $\varepsilon_F$ to the wavelet coefficients $F$, i.e., $\max|F-\check{F}|=\varepsilon_F$. 
The pointwise error in the physical space, $f-\check{f}$, is a weighted sum of $F-\check{F}$, with the weight determined by the lifting coefficients
and by the length of the filter. From this consideration, it may be possible to evaluate $\eta$ analytically, 
but the empirical value of $1.75$ proves acceptable in all test cases considered in this paper, see \ref{sec:error_control}.
We then assign $\delta^{\langle J \rangle} = \varepsilon_F$.
The values of $Q_i^{\langle j \rangle}$, $\delta^{\langle j \rangle}$, $F^{\langle j \rangle}_{\min}$ and $J$ are determined as follows from Algorithm~\ref{algo:quant},
where we use square brackets to denote the nearest integer.
An example graphical illustration of this algorithm is presented in figure~\ref{fig:quant}.
In the example, $J=3$ bit planes are required to represent the floating-point data with the desired accuracy $\varepsilon$.

\begin{algorithm}[H]
 \SetAlgoNoLine
 \SetInd{0.5em}{0.5em}
 \KwData{tolerance $\varepsilon$; point values $F_i$, $i=1,...,N$.}
 \KwResult{bit depth parameter $J$; offsets $F^{\langle j \rangle}_{\min}$; quantization steps $\delta^{\langle j \rangle}$; bit planes $Q_i^{\langle j \rangle}$, $i=1,...,N$, $j=1,...,J$.}
 $F_i^{\langle 1 \rangle} = F_i$, $i=1,...,N$\;
 $\varepsilon_F =  \varepsilon \max |f_{i_x,i_y,i_z}| / \eta$, $i_x=1,...,n_x, i_y=1,...,n_y, i_z=1,...,n_z$\;
 $q = 256$\;
 $j \gets 1$\;
 \Repeat{$\delta^{\langle j \rangle} < \varepsilon_F$}
   {
   $F^{\langle j \rangle}_{\min} = \min_{k=1,...,N} F_k^{\langle j \rangle}$\; 
   $F^{\langle j \rangle}_{\max} = \max_{k=1,...,N} F_k^{\langle j \rangle}$\;
   \eIf{$(F^{\langle j \rangle}_{\max} - F^{\langle j \rangle}_{\min})/(q-1) > \varepsilon_F$}{
    $\delta^{\langle j \rangle} = (F^{\langle j \rangle}_{\max} - F^{\langle j \rangle}_{\min})/(q-1)$\;
   }{
    $\delta^{\langle j \rangle} = \varepsilon_F$\;
   } 
   \For{$i=1$ \rm{to} $N$}{
    $Q^{\langle j \rangle}_i = \left[ (F^{\langle j \rangle}_i-F^{\langle j \rangle}_{\min})/\delta^{\langle j \rangle} \right]$\;
    $F^{\langle j+1 \rangle}_i = F^{\langle j \rangle}_i - (Q^{\langle j \rangle}_i \delta^{\langle j \rangle} + F^{\langle j \rangle}_{\min})$\;
   }
   $j \gets j+1$\; 
   }
 \caption{Quantization algorithm.}\label{algo:quant}
\end{algorithm} 
\vspace{5mm}

\subsection{\label{sec:entropycoding}Entropy coding}

Entropy coding is 
a technique allowing 
to reduce the quantity of storage required to hold a message
without any loss of information.
The message can be any sequence of characters drawn from some selected alphabet.
The implementation that we use in the present work is based on an alphabet of 256-bit symbols,
i.e., integers from 0 to 255. 
Conceptually, entropy coding consists in representing frequently
occurring sub-sequences with few bits and rarely occurring
ones with many bits.
Shannon's coding theorem serves the entropy of a message as 
a theoretical bound to possible lossless
compression \cite{Shannon_1948_bell}.

The entropy coding method that we employ in our present work is the range coding \cite{Martin_1979_conf}.
Our current implementation uses an open-source coder \textit{rngcod13} developed by 
M. Schindler \cite{Schindler_1999_website}.
The size of the encoded files produced by this coder 
is within a fraction of per cent from the theoretical bound,
and the operation speed is faster compared to other similar methods such as arithmetic coding.

The input message consists of the bit planes $Q_{i}^{\langle j \rangle}$ obtained during the quantization step.
These data arrays are divided in blocks of 60,000 elements. 
The frequencies of each value in a block are counted and copied to the output stream of the 
coder. The data block is then encoded using the calculated frequencies.
Given a stream of 256-bit symbols $Q_{i}^{\langle j \rangle}$ and their frequencies, the coder produces a shorter stream of bits to represent these symbols. The output stream is written in a file, appended with metadata necessary for decoding.

\section{Software framework}
\label{sec:software}

This section contains an overview of the software.
Installation instructions and user's notes can be found in 
{\tiny \url{https://github.com/pseudospectators/WaveRange/blob/master/README.md}}

\subsection{Software functionalities}
\label{}

WaveRange application includes an encoder that 
compresses the floating point data and a decoder that reconstructs the data from the compressed format.
Two different executables are built for these two purposes, respectively, when WaveRange is used as a standalone application.
WaveRange's `generic' interface can read and write
plain Fortran and C/C++ files containing one or several multi-dimensional arrays.
In addition, the current version of WaveRange includes `specialized'
input and output interfaces compatible with
a spectral incompressible Navier--Stokes solver FluSI \cite{Engels_2016_sisc} and
with MSSG, the Multi-Scale Simulator for the Geoenvironment \cite{Takahashi_2013_jop}.
Each of these two solvers generate two types of 
large files: regular output for, e.g., flow visualization, and restart files.
WaveRange can compress both.
The input floating point data for compression may be stored in one file or divided in multiple files,
in a format specific to the computation software employed.

WaveRange can also be used as a library. In that case, the encoder and the decoder functions
are called from the user's program. The original, the encoded and the reconstructed datasets are passed as parameters.
Disk input/output is to be implemented by the user.

\subsection{Software architecture}
\label{}

WaveRange is written in C and C++. 
The lower-level functions related with the discrete wavelet transform and range coding are in C. Their source codes are stored in the directories 
\verb|waveletcdf97_3d/| and \verb|rangecod/|, respectively.
On a higher level, the C++ code in \verb|core/| 
implements the functions \verb|void encoding_wrap| and \verb|void decoding_wrap|
that execute all operations necessary for compression and reconstruction, respectively,
in accordance with the data flow diagram in Fig.~\ref{fig:data_flow}.
These functions are called after the input data files are read and before the output files are written.
Two similar functions, \verb|void encoding_wrap_f| and \verb|void decoding_wrap_f|, offer compatibility with Fortran.
The \verb|int main| functions are specific to each interface. The respective C++ source files are 
contained in the directories \verb|generic/|, \verb|flusi/| and \verb|mssg/|, the executables are generated in \verb|bin/generic/|, \verb|bin/flusi/| and \verb|bin/mssg/|.
Note that the FluSI interface requires the hierarchical data format (HDF5) library \cite{Folk_1999_ps}, but the generic and MSSG interfaces 
have no external dependencies. The WaveRange library files appear in \verb|bin/lib/|.




\section{\label{sec:results}Illustrative examples}

In the subsequent sections of this paper, 
we discuss numerical examples that serve to 
evaluate the performance of the method and to gain
better understanding of different properties of the compressed data.
A homogeneous isotropic turbulence dataset, which can be regarded as a highly idealized representation of 
atmospheric flow, is examined in Section~\ref{sec:hit_compression}.
Quantitative measures such as the error norm, the relative compressed file size and 
the compression ratio are introduced, their scaling 
with respect to the tolerance and the data size is analyzed.
A similar analysis with application to seismology is discussed in Section~\ref{sec:seismology}.
After gaining insight into the basic typical features of the compressed data,
the discussion proceeds to realistic atmospheric flow simulations.
In Section~\ref{sec:tymip_restart}, a global weather simulation for typhoon prediction is considered,
with special attention to restart of the simulation from a compressed restart data file.
Restarts using different levels of compression tolerance are tested. 
Restarts of an urban-scale weather simulation are
discussed in Section~\ref{sec:bayarea}.

\subsection{\label{sec:hit_compression}Fluid turbulence simulation}

The homogeneous isotropic turbulence (HIT) dataset considered here is similar to the velocity fields analyzed in \cite{Onishi_2013_jcp}.
It was obtained by integrating the incompressible Navier--Stokes
equations in a $2\pi$-periodic cube box, in dimensionless units, using a
fourth-order finite-difference scheme, second-order Runge--Kutta time marching, and HSMAC velocity-pressure coupling \cite{Onishi_2011_jcp}. 
The flow domain was discretized using a uniform Cartesian grid consisting of $n^3 = 512^3$ staggered grid points.
Power input was supplied using external isotropic forcing 
to achieve a statistically-stationary state.
A snapshot velocity field ($u$,$v$,$w$) was stored in double precision,
each component occupying 1~GB of disk space.

\begin{figure}
    \begin{center}
        \includegraphics[width=0.483\textwidth,clip]{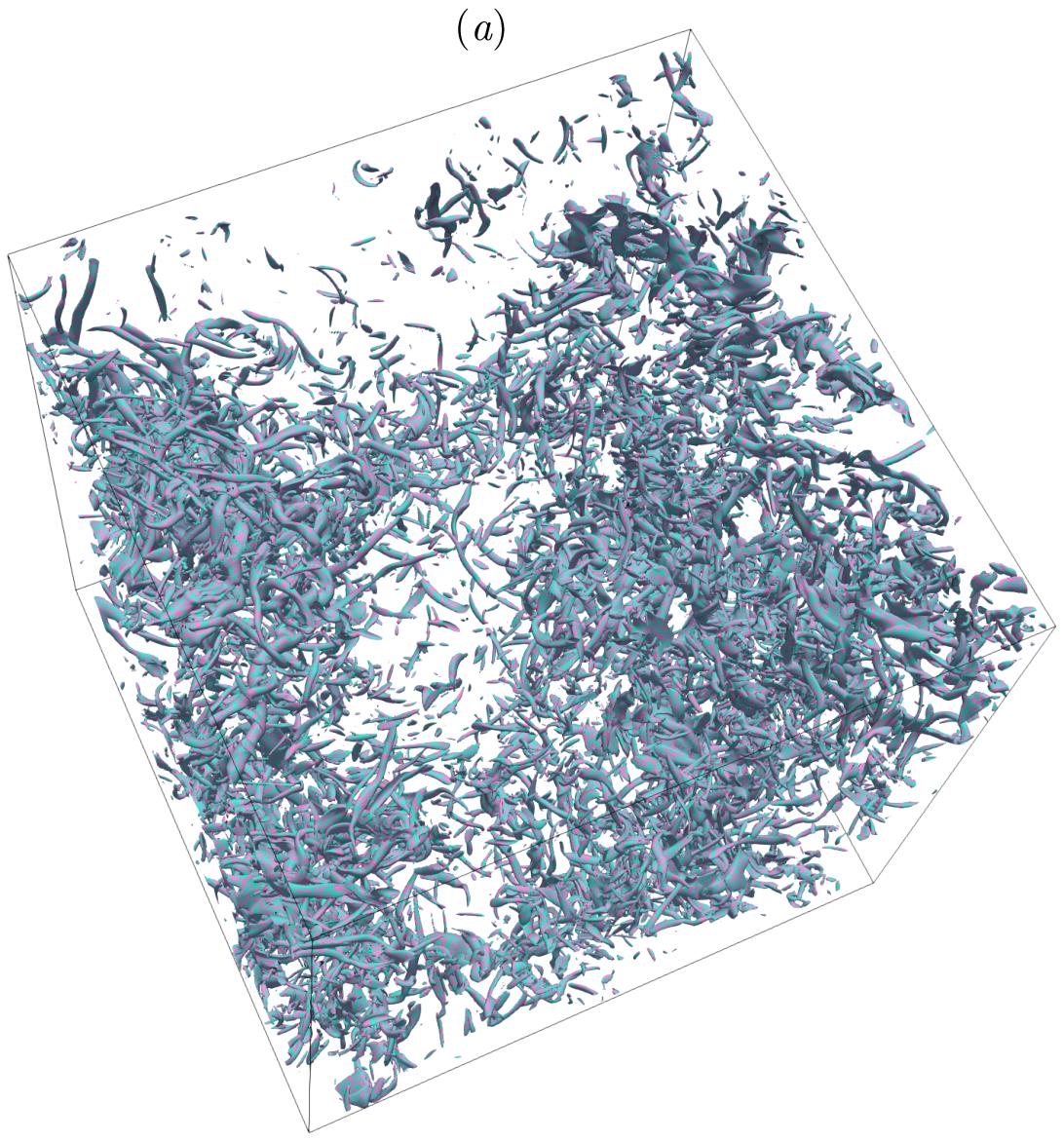} \quad
        \includegraphics[width=0.444\textwidth,clip]{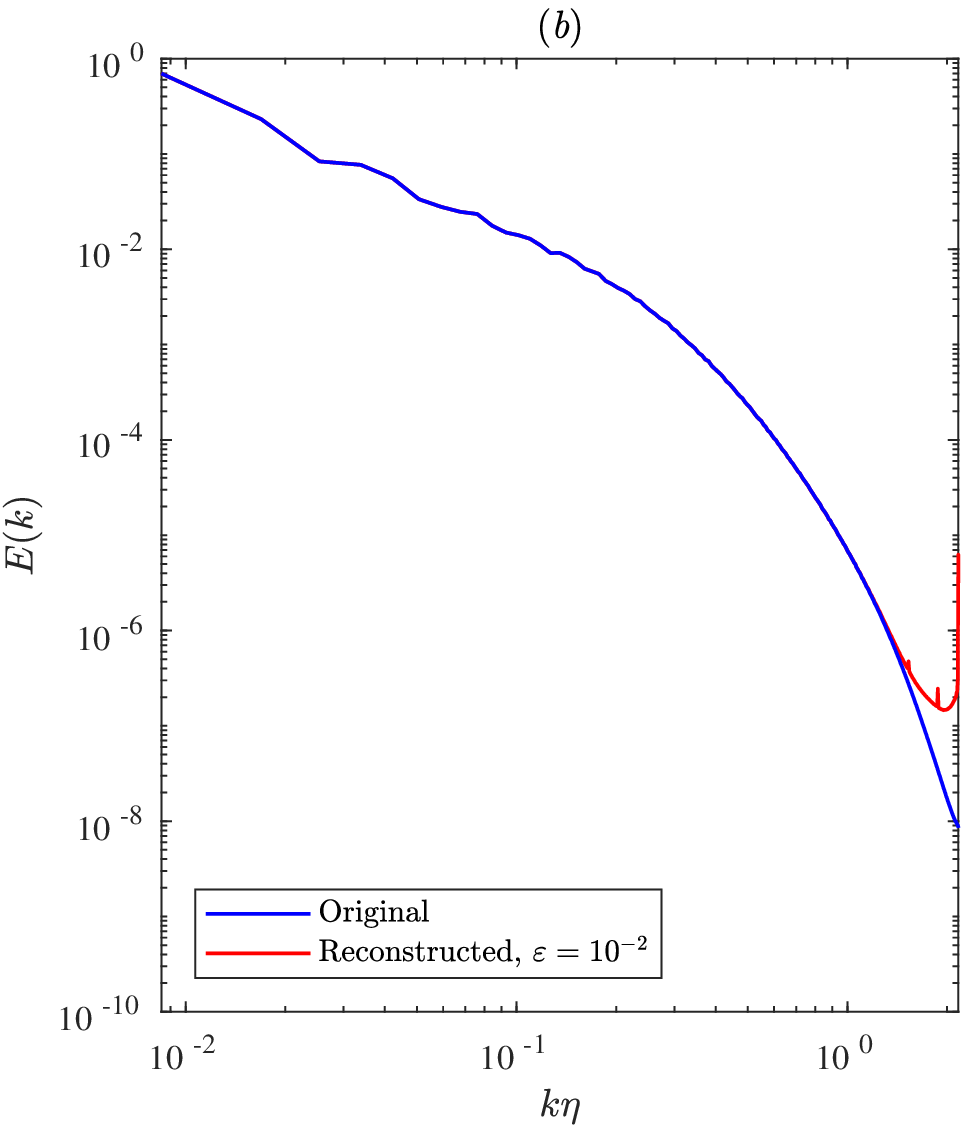}        
    \end{center}
    \caption{The HIT dataset. (\textit{a}) Flow visualization using a vorticity magnitude iso-surface 
    corresponding to 15\% of the maximum value. Black lines show the extent of the computational domain. Iso-surfaces of the
    original field (cyan) and reconstructed after compression with $\varepsilon=10^{-2}$ (magenta) 
    are superimposed and overlap almost perfectly. (\textit{b}) Energy spectra of the original and the reconstructed velocity fields.}
    \label{fig:hit}
    \vspace*{-3mm}
\end{figure}

The dataset used in the present study is visualized in Fig.~\ref{fig:hit}(\textit{a}) using 
an iso-surface of the vorticity magnitude, and the energy spectrum is shown in Fig.~\ref{fig:hit}(\textit{b}). 
Blue color corresponds to the original data.
The three velocity components are in the range $[-4.08, 4.16]$, $[-5.20, 4.82]$ and $[-4.31, 4.74]$, respectively.
The turbulence kinetic energy is equal to $K=1.44$.
With the dissipation rate $\epsilon=0.262$, the Kolmogorov length scale is equal to $\eta=(\nu^3/\epsilon)^{1/4}=0.00845$,
such that $k_{max}\eta = 2.16$, where $k_{max}=n/2$.
The Taylor micro scale $\lambda=\sqrt{10 \nu K / \epsilon}=0.246$ yields the Reynolds number $Re_\lambda=\sqrt{2K/3} \lambda / \nu=218$. 
Homogeneous isotropic turbulence presents a challenge for the data compression.
Turbulent flow fills the entire domain with fluctuations at all scales, 
the smallest scale being of the same order as the discretization grid step (see figure~\ref{fig:hit}\textit{b}).

\begin{figure}
    \begin{center}
        \includegraphics[scale=0.7]{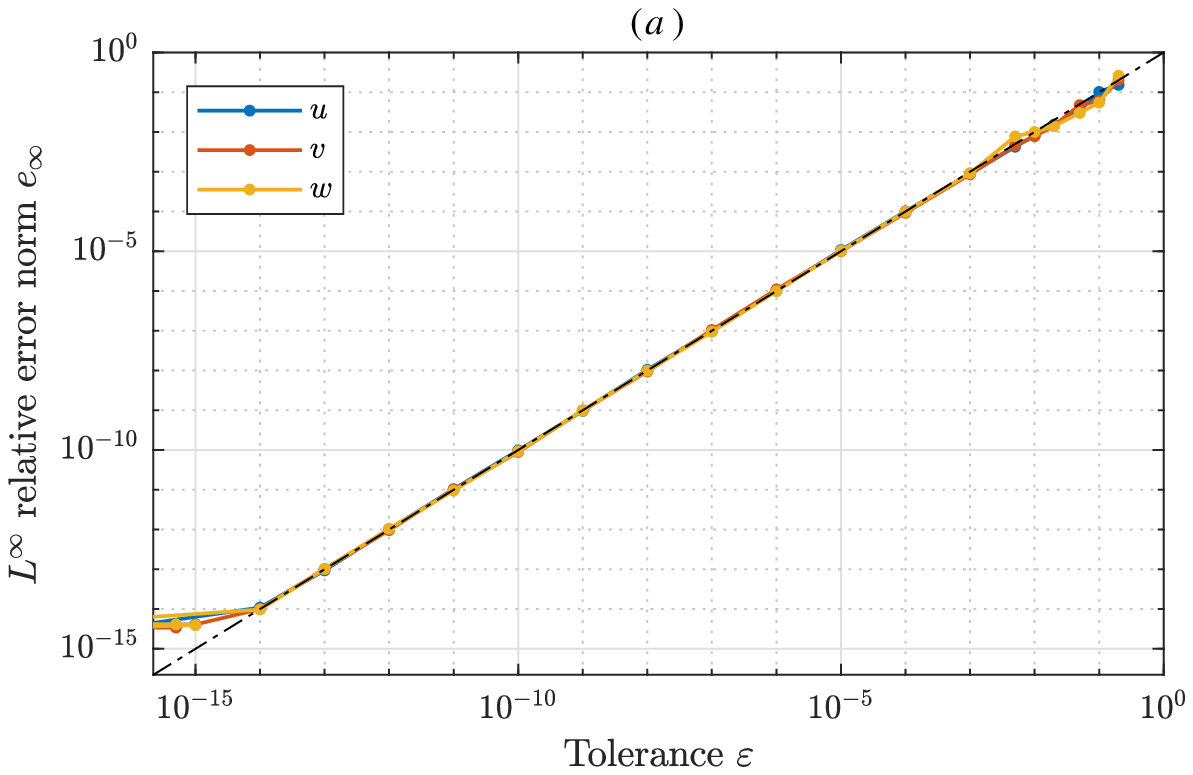}\includegraphics[scale=0.7]{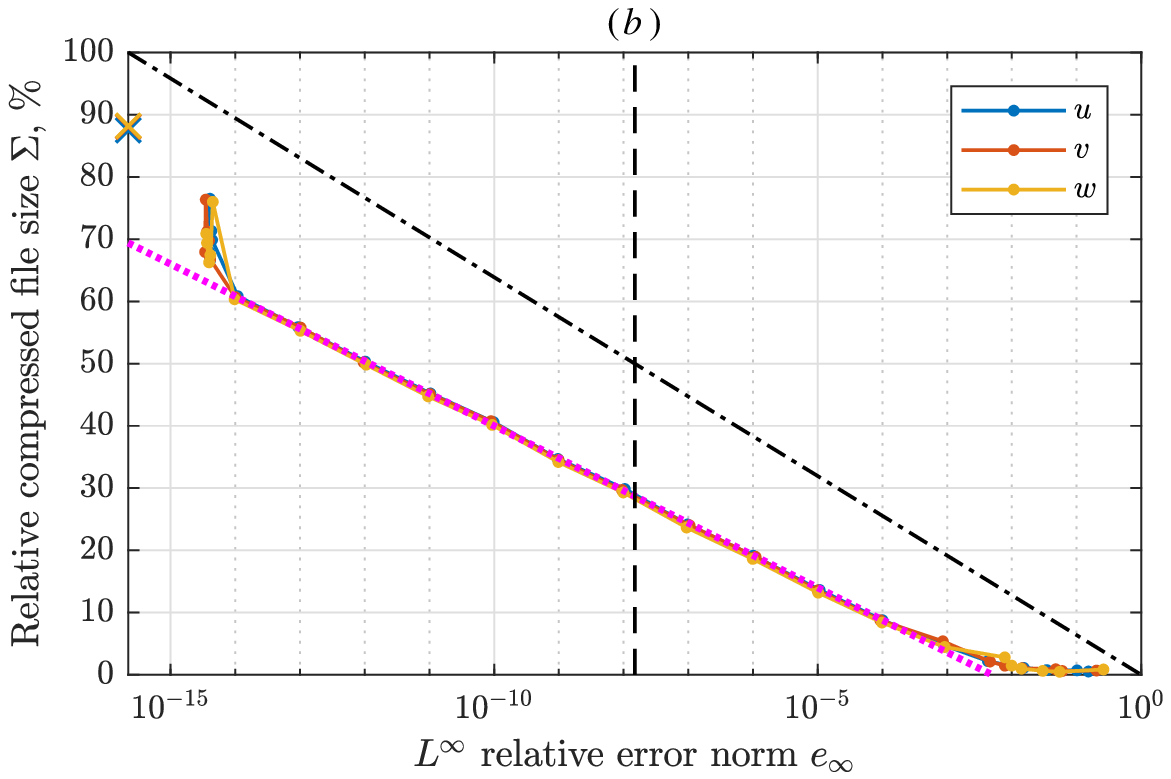}
        \includegraphics[scale=0.7]{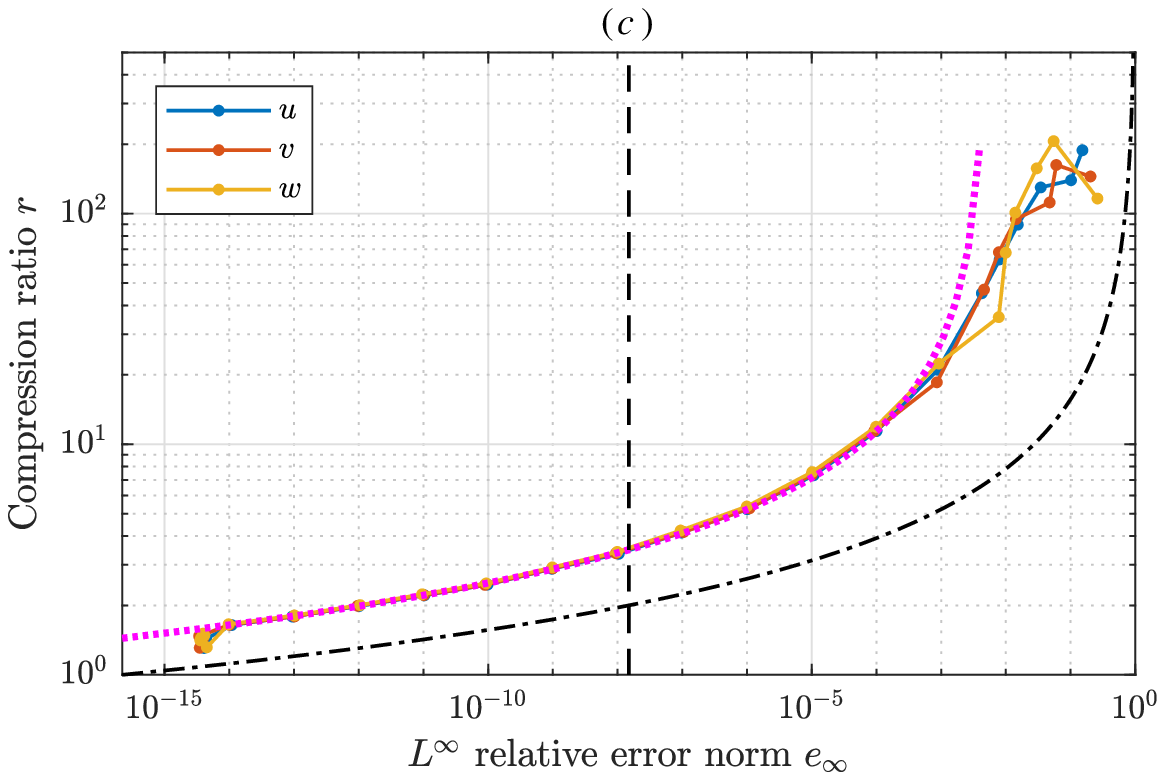}                
    \end{center}
    \caption{Compression of the HIT velocity components.
    (\textit{a}) $L^\infty$ error norm as a function of the tolerance $\varepsilon$;
    (\textit{b}) Compressed file size in per cent of the original file size, as a function of the $L^\infty$ error norm;
    (\textit{c}) Compression ratio versus relative $L^\infty$ error.
    The fit (\ref{eq:sigma_hit}) is shown with a dotted line.
    The dashed vertical line shows the accuracy of single-precision storage and
    the dash-dot lines correspond to compression using quantization only.}
    \label{fig:hit_res}
\end{figure}

Before analyzing the compression performance in this case, let us discuss the error control properties.
Among the variety of physically motivated error metrics we choose $L^\infty$, which is intuitive and perhaps the most stringent.
Other, problem specific metrics will be discussed later on.
Figure~\ref{fig:hit_res}(\textit{a}) shows the relative $L^\infty$ error of the velocity field 
reconstructed from compressed data. The compression algorithm takes the tolerance $\varepsilon$
as a control parameter. This allows to plot the reconstruction error as a function of $\varepsilon$.
Each velocity component is scaled by its maximum absolute value, yielding
\begin{equation}
e_\infty = \frac{ \max{ |\check{f} - f| } }{\max{|f|}},
\label{eq:errlinf}
\end{equation}
where $f$ stands for one of the components ($u$, $v$ or $w$) of the original velocity field,
and $\check{f}$ is its reconstruction from the compressed data. 
The maxima and the minima are calculated over all grid points.
In figure~\ref{fig:hit_res}(\textit{a}), the dash-dot diagonal line 
visualizes the identity relationship between $\varepsilon$ and $e_\infty$.
The actual data for all three velocity components closely follows this trend,
with the discrepancy only becoming noticeable when $\varepsilon$ is smaller than
the accumulated roundoff error, and for large $\varepsilon$ when
the discrete nature of quantization becomes apparent as 
there remain only few non-zero detail coefficients. 
For all $\varepsilon \in [10^{-14},10^{-3}]$, the difference between $\varepsilon$ and $e_\infty$ is less than 15\%,
i.e., the desired error control is successfully achieved.

Figure~\ref{fig:hit_res}(\textit{b}) shows a plot of the relative compressed file size 
\begin{equation}
\Sigma=\frac{s(\varphi)}{s(f)} \times 100\%
\label{eq:sigma}
\end{equation}
versus $e_\infty$, where
$s(f)=1$~GB is the disk space required to store a $512^3$ array of double-precision values,
and 
$s(\varphi)$ is the respective compressed file size in GB.
The dash-dot diagonal line in this plot represents the storage requirement for a
$512^3$ array of hypothetical custom-precision values, which is 100\% in the case of double precision,
50\% for the single precision, 0 for the full loss of information, and all intermediate values are linearly interpolated. 
This line sets a reference it terms of the compression achievable by simply quantizing the 
floating-point array elements with tolerance $\varepsilon$ and using smaller, but constant number of bits per element.
Its slope is equal to 6.39\% of storage per one decimal order of magnitude of accuracy.

The point markers connected by solid lines show the 
amount disk space actually used by the compressed velocity fields files,
and the respective reconstruction error.
When the accuracy is in the range between $10^{-14}$ and $10^{-3}$,
it is well approximated with a fitting
\begin{equation}
\Sigma_{HIT} = (-0.12 - 0.052 \log_{10}{e_\infty}) \times 100\%,
\label{eq:sigma_hit}
\end{equation}
which is shown with a dotted magenta line.
The negative intercept can be explained as follows.
When $\varepsilon$ (and, consequently, $e_\infty$) is large,
only a few significant bits are sufficient to represent the field $(u,v,w)$
with the desired accuracy. Hence,
the wavelet transform after quantization becomes sparse,
and it is very efficiently compressed by entropy coding.
In the intermediate range of $\varepsilon$,
increasing accuracy by one order of magnitude comes at a cost of $5.2\%$ increased storage.
Least significant bit planes of the wavelet coefficients are not sparse, they are noise-like,
and their lossless compression ratio by entropy coding asymptotically tends to 
a fairly low value typical of noise, $0.0639/0.052 \approx 1.23$, in the hypothetical limit of $\varepsilon \to 0$.
However, for $\varepsilon<10^{-14}$, accumulation of roundoff errors becomes significant,
$\Sigma$ sharply increases to 76\%, while $e_\infty$ saturates at $5 \times 10^{-15}$.
Roundoff error could be avoided by switching to integer wavelet transform 
ensuring perfect reconstruction. 
Extrapolation of the linear trend suggests that 
this approach may be superior than, for example, direct application of LZMA encoding, as
shown by crosses in figure~\ref{fig:hit_res}(\textit{b}).
It is also noteworthy that compression with $\varepsilon=10^{-8}$ allows to reduce the data storage by a factor of 3, which is significantly better than 
using the native single-precision floating-point format.
Compression with $\varepsilon=10^{-6}$ increases this ratio to 5. 
On the other hand, 
from the figure it is apparent that multi-fold reduction
of the file size, which is our objective, entails data loss. 

Compression ratio 
\begin{equation}
r=\frac{s(f)}{s(\varphi)}
\label{eq:crt}
\end{equation}
is a commonly used performance metrics, particularly
suitable in those situations when the compressed file size is many times smaller 
than the original size.
Thus, Figure~\ref{fig:hit_res}(\textit{c}) reveals that hundredfold compression is attainable 
by setting $\varepsilon=10^{-2}$. This may be a good setting for the purpose of qualitative flow visualization, for example,
as illustrated in figure~\ref{fig:hit}(\textit{a}): vortex filaments consist of many small cyan and magenta patches,
which means that two iso-surfaces coincide almost perfectly, namely,
the cyan iso-surface that visualizes the vorticity magnitude calculated from the original velocity,
and the magenta iso-surface that is obtained using the velocity field reconstructed from the compressed data with $\varepsilon=10^{-2}$.
Figure~\ref{fig:hit}(\textit{b}) confirms that the error mainly contaminates the smallest scales,
while the energy-containing large-scales are much less affected.
This observation supports the interpretation of quantization as adding thermal noise to the original data.
Note that we chose $\varepsilon=10^{-2}$ to magnify the numerical error. For $\varepsilon = 10^{-6}$ or less, the reconstructed spectrum would be visually identical
with the original.
Finally, figure~\ref{fig:hit_res}(\textit{c}) confirms that asymptotic scaling (\ref{eq:sigma_hit}) is realized 
as soon as $e_\infty < 10^{-3}$.

\begin{figure}
    \begin{center}
        \includegraphics[scale=0.7]{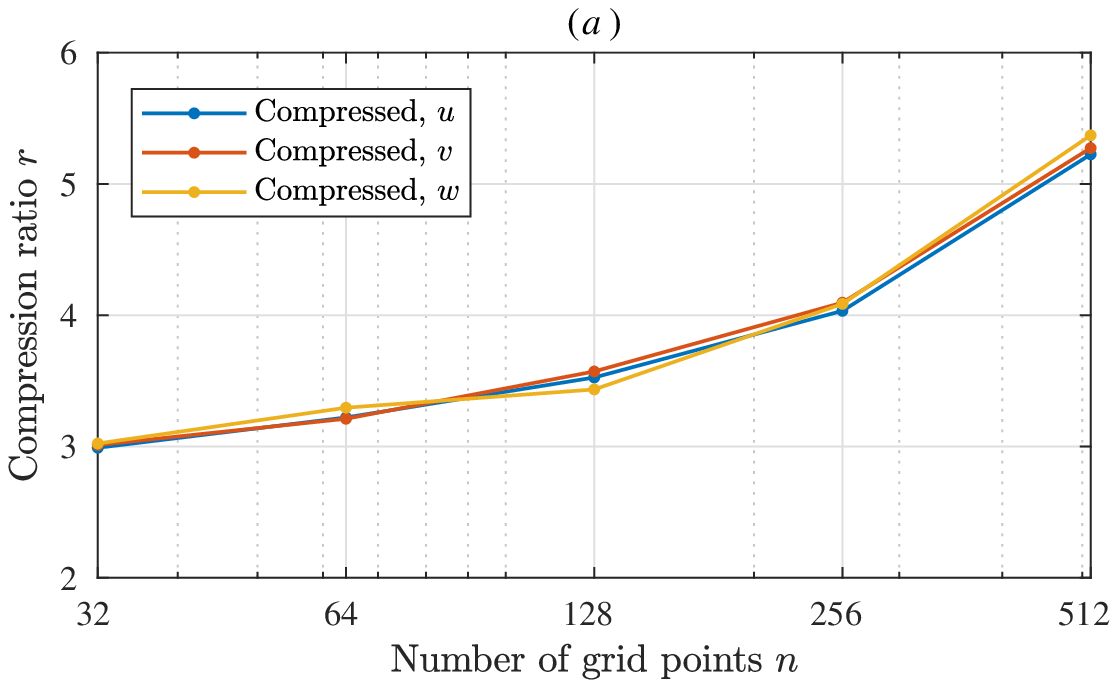}\includegraphics[scale=0.7]{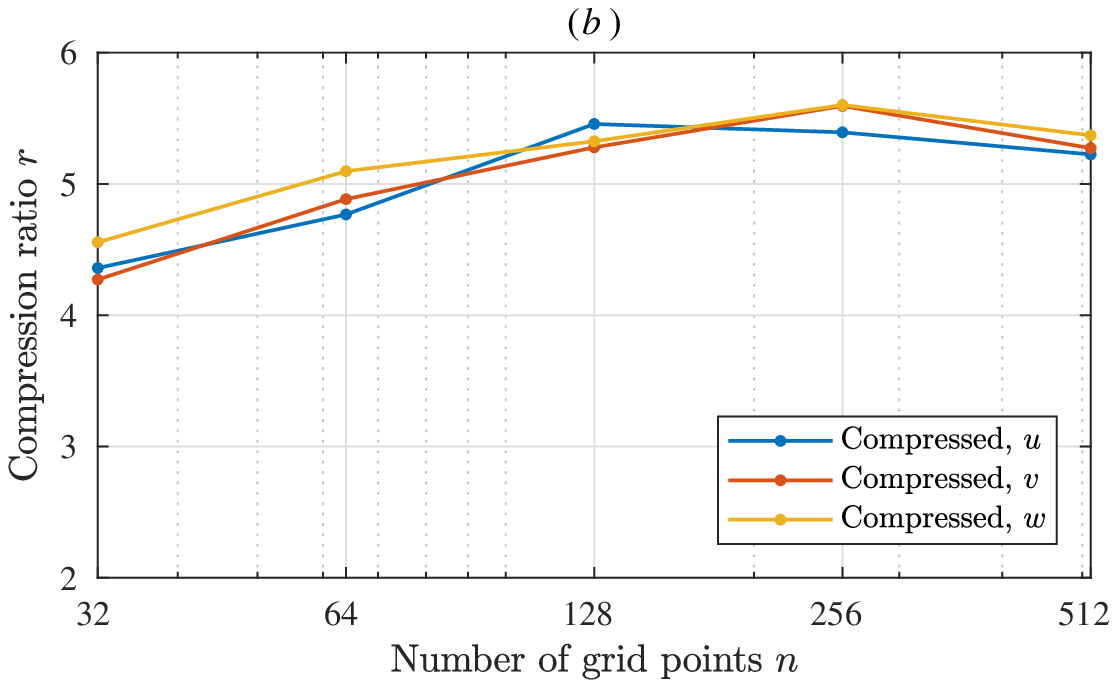}                
    \end{center}
    \caption{Compression ratio $r$ of (\textit{a}) uniformly down-sampled velocity datasets
    containing $n^3$ out of $512^3$ grid points;
    (\textit{b}) velocity data in sub-domains of size $n^3$.}
    \label{fig:hit_extract}
\end{figure}

Let us now consider how the compression ratio $r$ scales with the size of the dataset.
There are at least two obvious ways to obtain a smaller HIT dataset from
the original $512^3$ arrays: down-sampling and sub-domain extraction. 
The former means that only every 2nd (or 4th, etc.) grid point in each direction is retained, 
all other data points are discarded. The latter means that only the first 256 (or 128, etc)
points in each direction are retained. We have thus constructed two sequences of datasets of
size $n^3=32^3$, $64^3$, $128^3$, $256^3$ and $512^3$, where the largest dataset is the original one.
The compression ratio using tolerance $\varepsilon=10^{-6}$ is displayed in figures~\ref{fig:hit_extract}(\textit{a})~and~(\textit{b}). 

Figure~\ref{fig:hit_extract}(\textit{a}) shows $r$ of the down-sampled data.
Starting from the rightmost point, there is a major decrease in $r$ between $n=512$ and $256$, i.e., after discarding every second point.
It is followed by a
slower decay that accompanies subsequent down-sampling.
The original dataset is a result of direct numerical simulation of turbulence,
which implies that the inertial range is fully resolved, i.e., the distance between
the neighboring grid points is smaller than the 
Kolmogorov scale. Consequently, the turbulent velocity fluctuation at the smallest scale
contains disproportionately less energy than any larger scale in the inertial range.
This means that the numerical values of the smallest-scale wavelet coefficients
contain much less non-zero significant digits.
Therefore, they are compressed much more efficiently. 
In the inertial range, the number of non-zero significant digits becomes larger as the wavenumber decreases,
which explains further gradual decrease of $r$ with decreasing $n$.
The evolution of $r$
with $n$ is thus related to the decay of the Fourier energy spectrum.

Figure~\ref{fig:hit_extract}(\textit{b}) shows $r$ of the sub-domain data.
In this case, 
the energy per unit volume of the smallest-scale velocity fluctuation field does not depend on $n$.
Hence, smallest-scale wavelet coefficients are of the same order of magnitude, regardless of $n$.
As a result, in this case, the compression ratio $r$ varies less with $n$ than in the previous case.
It is practically constant, $r=5.5$, between $n=128$ and $512$.
There is, nevertheless, some moderate decrease
down to $r=4.5$ at $n=32$. It can be explained by the efficiency
of the entropy coding becoming lower as the dataset becomes smaller.

These scalings confirm that the chosen method of compression is particularly suitable for datasets
produced by large-scale, high-resolution numerical simulations.
In addition, the largest compression ratio will be achieved if the data are stored in one single file 
rather than divided in multiple sub-domain files treated independently.
Further, \ref{sec:wake_compression} investigates into the effects of spatial inhomogeneity 
by considering fluid flow past a solid cylinder.

\subsection{\label{sec:seismology}Seismology simulation}

\begin{figure}
    \begin{center}
        \includegraphics[scale=0.7]{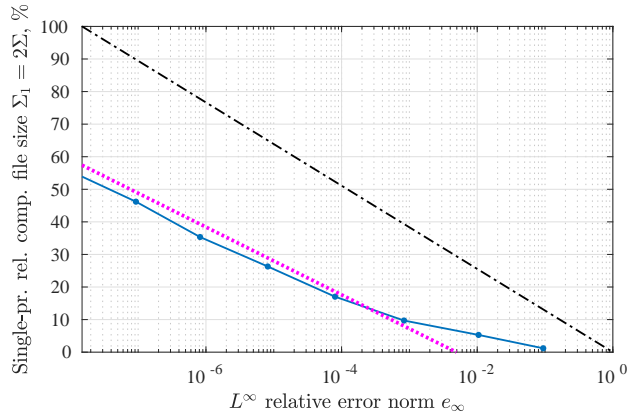}
    \end{center}
    \caption{Compression of the seismology simulation restart file.
    Compressed file size in per cent of the original file size, as a function of the $L^\infty$ error norm;
    The fit reference (\ref{eq:sigma_hit}) is shown with a dotted line.
    The dash-dot line corresponds to compression using quantization only.}
    \label{fig:dms_res}
\end{figure}

Let us proceed with an example from seismology. Similarly to the analysis in the previous section, we compress an example dataset with a given $\varepsilon$, measure the compressed file size, then reconstruct the fields from the compressed format and measure the $L^\infty$ error $e_\infty$.
The example dataset is a restart file for a synthetic seismogram computation of a large earthquake using SPECFEM3D, a spectral-element code based on a realistic fully three-dimensional Earth model \cite{Komatitsch_2013_book}.
In the simulation, the domain is split in as many slices as the number of processors used. Data that correspond to different slices are stored in separate files,
and, in this example, we only process one of them. The file contains 31 single-precision arrays, but only the first 9 are large: 
the displacement, velocity and acceleration of the crust and mantle ($3 \times 19904633$ elements each),
of the inner core ($3 \times 1270129$ elements each) and of the outer core ($2577633$ elements each).
Only these arrays are compressed, while the remaining 22 small arrays containing in total $5250$ single-precision numbers are directly copied in the end of the compressed file. Note that, although the grid is three-dimensional, the spectral element solver packs the variables in one-dimensional arrays. This packing preserves spatial 
localization, therefore, wavelet compression remains efficient. Figure~\ref{fig:dms_res} shows the
relative compressed file size $\Sigma_1=s(\varphi)/s(f_1)$ versus $e_\infty$, where $s(f_1)=757$~MB is the original single-precision file size.
The result is close to the trend obtained in the previous sections for the turbulent fluid flow, the latter shown with a dotted line.

\subsection{\label{sec:tymip_restart}Global-scale simulation of tropical cyclones}

In this section, we evaluate the compression of restart files produced in a global weather simulation using the
Multi-Scale Simulator for the Geoenvironment (MSSG), which is a coupled non-hydrostatic atmosphere-ocean-land model
developed at the Center for Earth Information Science and Technology, Japan Agency for Marine-Earth Science and Technology (JAMSTEC) 
\cite{Takahashi_2013_jop}. Its atmospheric component 
includes a Large-Eddy Simulation (LES) model for the turbulent atmospheric boundary layer and a six-category
bulk cloud micro-physics model \cite{Onishi_2012_jas}. Longwave and shortwave radiation transfer is
taken into account using the Model simulation radiation transfer code version 10 (MstranX) \cite{Sekiguchi_2008_jqsrt}.
In the global weather simulation mode, MSSG uses a Yin-Yang grid system in order to relax Courant-Friedrichs-Lewy condition on the sphere.
Higher-order space and time discretization schemes are employed. 
 
In \cite{Nakano_2017_gmd}, nine tropical cyclones were simulated
within the framework of the Global 7km mesh nonhydrostatic Model Intercomparison Project for improving TYphoon forecast (TYMIP-G7).
The main objective of that project was to understand and statistically quantify the advantages of high-resolution global atmospheric models towards the improvement of TC track and intensity forecasts.
Thus, in MSSG, each horizontal computational domain covered $4056 \times 1352$ grids in the Yin-Yang latitude-longitude grid  
system. The average horizontal grid spacing was 7~km. The vertical level comprised 55 vertical layers with a top height of
40~km and the lowermost vertical layer at 75~m. 
It was recognized that, when requiring the output data for every 1 or 3~h over 5-day periods be stored for analyses,
the total volume of storage summed up to a huge amount.

Here, we use the initial data of July 29, 2014 at 12:00~UTC.
The dataset is stored in multiple files.
They include a text header file that contains descriptive parameters of the dataset 
such as its size, hydrodynamic field identifiers, domain decomposition parameters, etc. 
The hydrodynamic field variables are stored in 1024 binary files,
each containing all $n_f=12$ fields within the same sub-domain. 
Thus, each sub-domain contains $n_f \times 254 \times 43 \times 55$ grid point data
in double precision, totaling to 55~MB of data in one file.
First 512 of these files belong to the Yin grid and the remaining 512 to the Yang grid.

The compression can either be performed on each data subset file independently
or, alternatively, the subsets can be merged before applying the wavelet transform.
It is noticed in Section~\ref{sec:hit_compression} that 
the compression ratio has a tendency to increase with the data size.
Hence, all subsets of each hydrodynamic field are concatenated as parts
of a three-dimensional array of size $4064 \times 2752 \times 55$.
The fields are processed sequentially requiring 4.6~GB of RAM 
for the input array plus up to 6.4~GB for the encoded output data and for
temporary arrays.

\begin{figure}
    \begin{center}
        \includegraphics[scale=0.7]{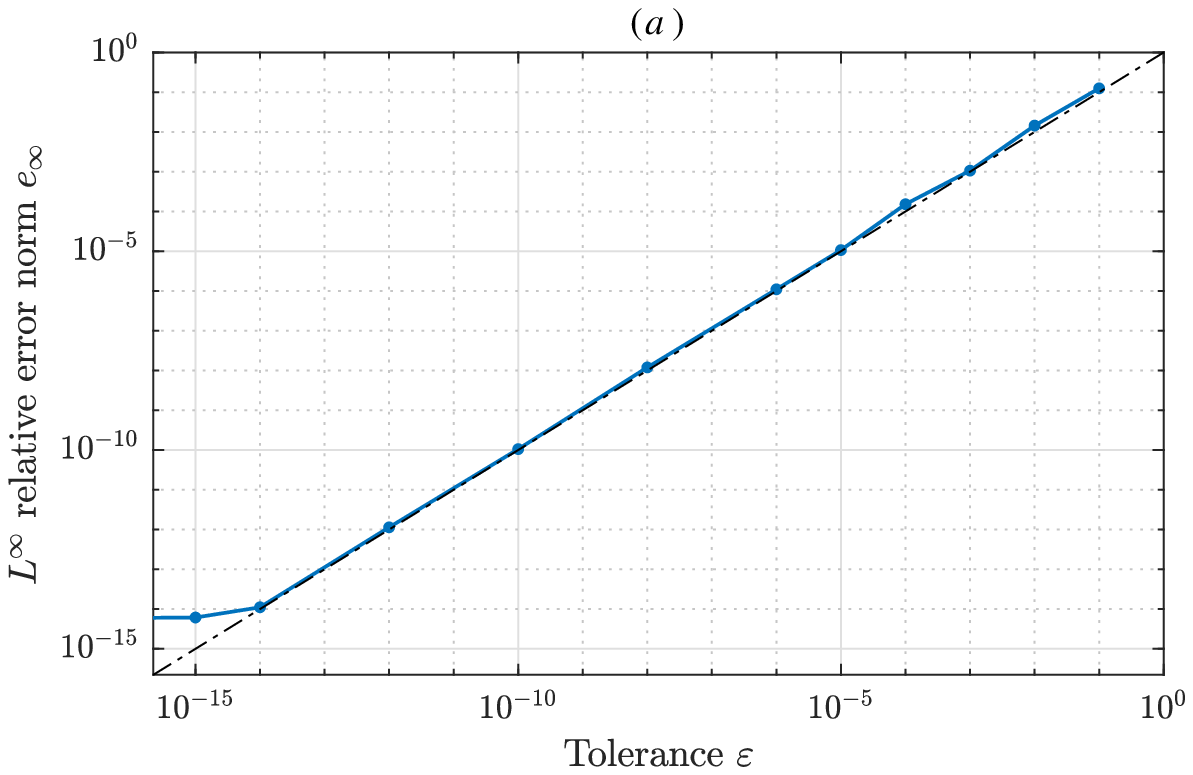}\includegraphics[scale=0.7]{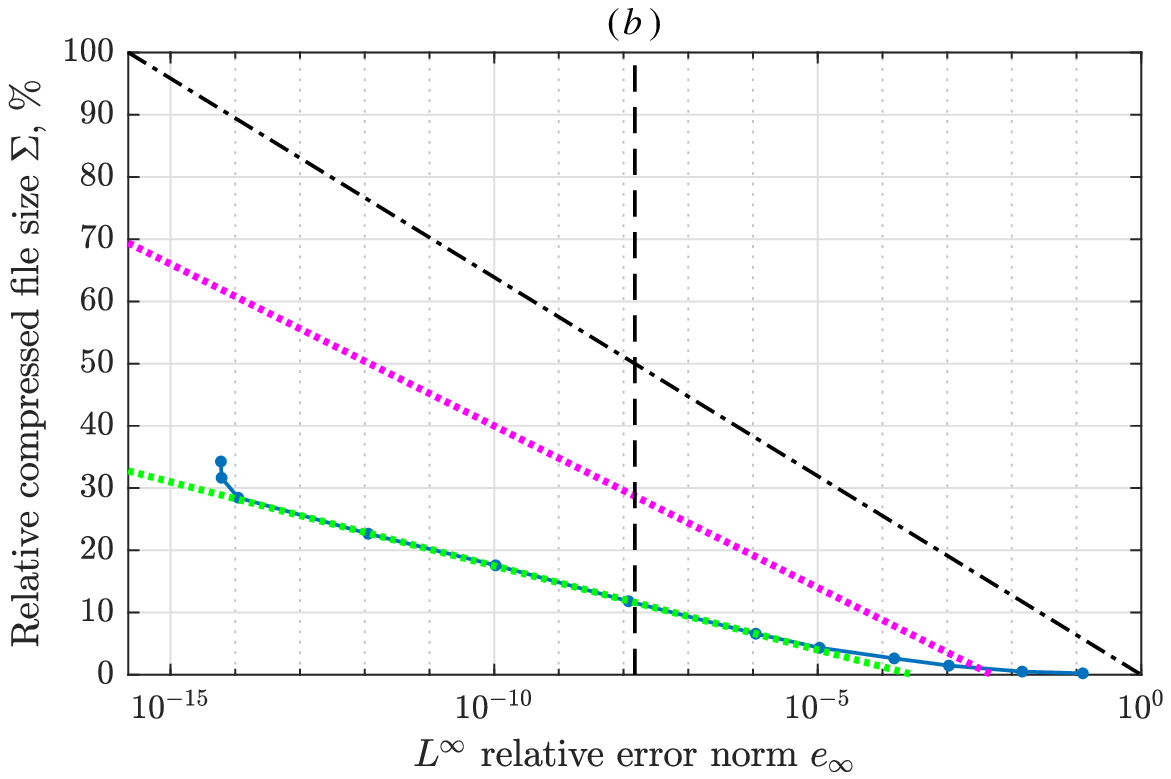}               
    \end{center}
    \caption{Compression of the typhoon simulation restart file.
    (\textit{a}) $L^\infty$ error norm as a function of the tolerance $\varepsilon$;
    (\textit{b}) Compressed file size in per cent of the original file size, as a function of the $L^\infty$ error norm.
    The fits (\ref{eq:sigma_hit}) and (\ref{eq:sigma_ty}) that correspond to $\Sigma_{HIT}$ and $\Sigma_{T}$, respectively, 
    are shown with the magenta and green dotted lines.
    The dashed vertical line shows the accuracy of single-precision storage and
    the dash-dot diagonal line corresponds to compression using quantization only.}
    \label{fig:tymip_compchart}
\end{figure}

First, let us
quantify the compression performance
of this restart dataset.
Figure~\ref{fig:tymip_compchart}(\textit{a}) shows that
the relative error $e_\infty$ varies almost identically to the threshold $\varepsilon$,
where $e_\infty$ is calculated as the maximum relative error over
all data points of all fields.
It saturates at the level of $10^{-14}$ due to round-off.    
The compressed file size, shown in figure~\ref{fig:tymip_compchart}(\textit{b}), is
significantly smaller than in the previously considered cases of turbulent incompressible velocity data.
The linear fit 
\begin{equation}
\Sigma_{T} = (-9.5 - 2.7 \log_{10}{e_\infty}) \times 100\%,
\label{eq:sigma_ty}
\end{equation}
shown with a green dotted line, has a greater offset from the dash-dot diagonal and a less steep slope 
compared with the HIT fit (\ref{eq:sigma_hit}), which is shown with a magenta dotted line. 
The global weather simulation is more complex than the incompressible Navier--Stokes,
the fields contained in the restart files are heterogeneous and have sparser wavelet transform that
compresses more efficiently. 

\begin{figure}
    \begin{center}
        \includegraphics[scale=1]{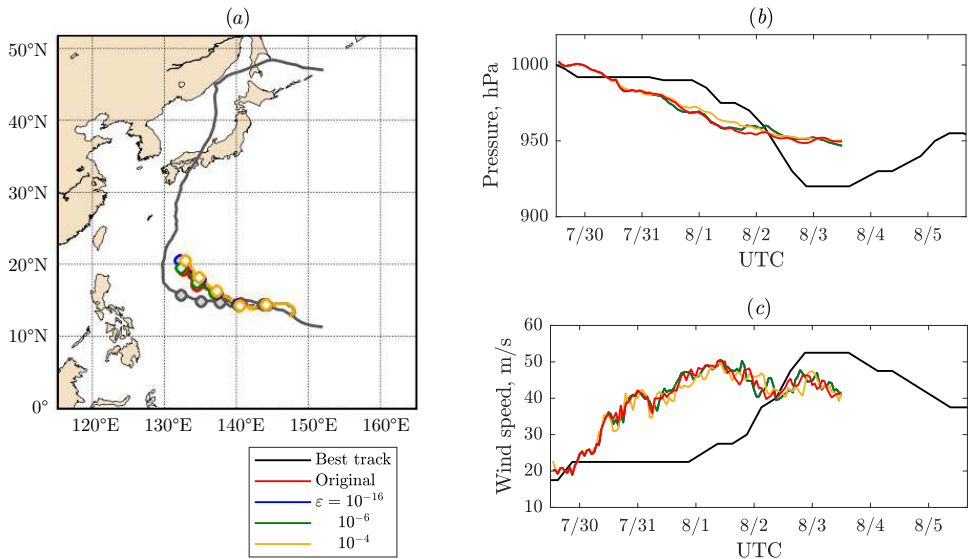}                   
    \end{center}
    \caption{Results of the typhoon forecast depending on the restart file accuracy $\varepsilon$:
    (\textit{a}) trajectory of the center; (\textit{b}) time evolution of the minimum pressure; (\textit{c}) time evolution of the maximum wind speed.}
    \label{fig:tymip_tracks}
\end{figure}

To measure the effect of lossy restart data compression on the simulation accuracy, the following protocol was implemented.
\begin{itemize}
\item Compress the original restart data with some given tolerance $\varepsilon$;
\item Using the compressed file, reconstruct the full-size restart data;
\item Restart the weather simulation and let it continue for 120 hours of physical time;
\item Compare the time evolution of selected physical quantities with the original simulation not using data compression.
\end{itemize}
We focus on typhoon Halong.
Figures~\ref{fig:tymip_tracks}(\textit{a}), (\textit{b}) and (\textit{c}) show
the typhoon core trajectory, minimum pressure in the core and the wind speed, respectively.
The best track from observation 
is shown using the black color, the result of the original simulation 
is shown using the red color.
The typhoon trajectory is predicted well during the first three days, after that it deviates more 
to the north during the simulation. 
The predicted pressure drop and the increase of wind speed are slightly advanced in time compared with the 
observation data, but the maximum wind speed is evaluated accurately.

The results of the restarts with $\varepsilon=10^{-16}$, 
$10^{-6}$ 
and $10^{-4}$
are shown with different colors.
All of them visually coincide with the original result during the first 24 hours,
after which the discrepancy grows large enough to be visible, but it remains in all cases 
significantly smaller than the difference with respect to the observation.

For $\varepsilon = 10^{-3}$ or larger, it was found impossible to restart the simulation because of numerical instability.
In fact, the onset of such instability is already noticeable in the case of $\varepsilon = 10^{-4}$, as
the wind speed becomes slightly different in the very beginning of that simulation.
We investigate further on this effect by plotting the difference between the restart and the original 
results on a logarithmic scale.
We consider the $L^2$ error norm of the wind speed, obtained by summation over all grid points in latitude-longitude square window $\Omega$ of 
size $10^\circ \times 10^\circ$ centered on the typhoon as predicted in the original simulation,
\begin{equation}
||U_{restart}-U_{original}||_2 = \left( \frac{1}{\#\Omega} \sum_{p \in \Omega} ( U_{restart}-U_{original} )^2 \right)^{1/2},
\label{eq:wind_l2err}
\end{equation} 
where $U_{restart}$ and $U_{original}$ is the velocity magnitude in the restart simulation and 
in the original simulation, respectively.
For all simulations with $\varepsilon \le 10^{-5}$, the error increases
polynomially as the power $\approx 2$ of the physical time after the restart point,
until saturation after about 72 hours.
This trend arises from the nonlinear dynamics of the system and it shows no distinguishable
deterministic relation with $\varepsilon$ as long as the latter is sufficiently small.
For $\varepsilon = 10^{-4}$, the error increases rapidly during the first time iterations,
but then it decreases and ultimately follows the same trend as described previously.
It is apparent that larger values of $\varepsilon$ entail faster
initial error growth and, ultimately, numerical divergence that cannot be accommodated by the 
physical model.

\begin{figure}
    \begin{center}
        \includegraphics[scale=0.7]{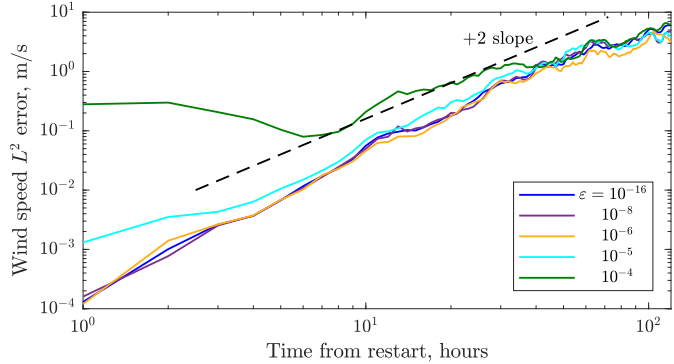}                      
    \end{center}
    \caption{Time evolution of the $L^2$ average error in the wind speed integrated over a $10^\circ \times 10^\circ$ window,
    for different levels of the restart file accuracy $\varepsilon$.}
    \label{fig:tymip_l2err}
\end{figure}

From figure~\ref{fig:tymip_compchart} we notice that successfull restarts belong to
the intermediate regime (where $\Sigma$ evolves according to equation (\ref{eq:sigma_ty}))
and to the high-accuracy tail of the compression diagram, such that
the compressed file size is greater than 2.5, i.e., the compression ratio is less than 40.
The low-accuracy end of the diagram showing the file size of less than 1\% can be practical for the data archiving
only if it is not intended as input for restarting the simulation.
Taking into consideration these opposing requirements of simulation reliability and data storage efficiency,
it is advisable to compress the restart data of global weather simulations with the relative tolerance of $\varepsilon=10^{-6}$.

\subsection{\label{sec:bayarea}Urban-scale simulation}

In \cite{Matsuda_2018_jweia}, a tree-crown resolving large-eddy simulation coupled with a three-dimensional radiative transfer model
was applied to an urban area around the Tokyo Bay. 
Source terms that represent contributions of the ground surface, buildings, tree crowns, and anthropogenic heat, were
integrated within the MSSG model in order to perform urban-scale simulations.
In particular, tree crowns were taken into account using the volumetric radiosity method.
The landscape was set based on geographic information system (GIS) data from the Tokyo Metropolitan Government.
The initial and side-boundary atmospheric conditions were imposed by the linear interpolation of the mesoscale data provided by the Japan Meteorological Agency.
The computational domain was a rectangular box discretized with uniform grid step (5~m) in two perpendicular horizontal directions, 
and slightly stretched in the vertical direction (from 5~m near the sea level to 15~m in the upper layers).
We consider restart data for a simulation using $N = 2500 \times 2800 \times 151$ grid points. 
Only the hydrodynamic fields are compressed since all other restart data use much less disk space.
The domain is
decomposed in equal Cartesian blocks of $50 \times 25 \times 151$ points. These data sets are stored in 5600 files, 
each containing 21 hydrodynamic fields. The files occupy 166~GB of disk space in total.

\begin{figure}
    \begin{center}
        \includegraphics[scale=0.7]{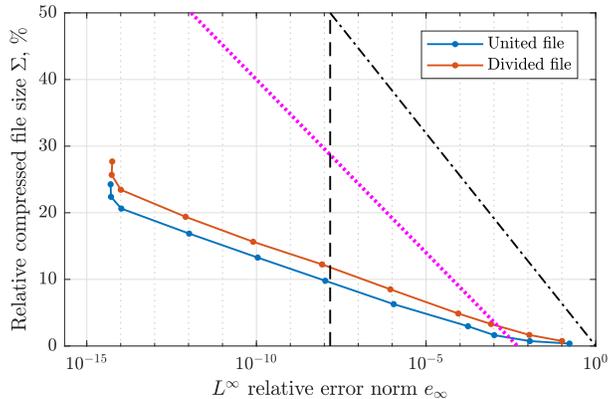}
    \end{center}
    \caption{Compression of the urban weather simulation restart file. Compressed file size in per cent of the original file size, as a function of the $L^\infty$ error norm.
    For reference, the dotted magenta line shows $\Sigma_{HIT}$ (\ref{eq:sigma_hit}), the dashed vertical line shows the accuracy of single-precision storage and
    the dash-dot diagonal line corresponds to compression using quantization only.}
    \label{fig:compression_urb}
\end{figure}

As shown in figure~\ref{fig:compression_urb}, data compression dramatically reduces the storage requirement. We have 
compared two approaches. The first (``united file") is to read the data from all sub-domains and concatenate
in one array per field, then transform and encode each field, and write all encoded data in one binary file.
This is the same procedure as used in Section~\ref{sec:tymip_restart}.
The second approach (``divided file") consists in processing each sub-domain 
independently, producing 5600 compressed binary files. 
Since it does not require any communication between sub-domains, processing can be executed in parallel with
ideal speedup. The parallel speedup comes at a price of larger compressed file size, by $\approx 2 \%$ of the original data size.
The sub-domain files are smaller than the united file, therefore, their compression ratio is overall lower,
as explained in Section~\ref{sec:hit_compression}.
In addition, part of the difference is due to the subdomain data being normalized with the respective
local maximum absolute value instead of using the global maximum.
The difference may be insignificant when considering the high-accuracy end of the plot, e.g., for $\varepsilon=10^{-14}$,
$\Sigma$ increases from 21 to 23\%. However, when the tolerance is set to $\varepsilon=10^{-6}$, 
the variation of the compressed file size from 6 to 8\% of the original restart file size can be considered as relatively large.
Using MPI communication for parallel wavelet transform and range coding, it may be possible to achieve better trade-off between parallel speedup and compression ratio.

We follow similar procedure as in Section~\ref{sec:tymip_restart} to
evaluate the effect of lossy compression upon restart.
Taking restart files from a previous simulation as initial data,
three new simulations have been performed 
for the period of 16:00-16:10 JST (Japan Standard Time) on August 11, 2007.
The first of them resumes from the original files,
while the second and the third resume from compressed data with $\varepsilon=10^{-6}$ and $10^{-12}$, respectively.
We use the united compressed file format that introduces no artifacts at the boundaries between subdomains.
In the following discussion, we analyze the output data written on disk in the end of these simulations.

Table~\ref{table_bay_err} shows the residual relative error in the $L^\infty$ and $L^2$ norms,
respectively calculated as
\begin{equation}
e_\infty(\varepsilon) = \frac{ \max_{i=1,...,N}{ |\check{f}_i(\varepsilon) - f_i| } }{\max_{i=1,...,N}{|f_i|}}
\quad\quad \textrm{and} \quad\quad e_2(\varepsilon) = \frac{ \left[ \frac{1}{N}\sum_{i=1}^N{ \left( \check{f}_i(\varepsilon) - f_i \right)^2 }\right]^{1/2} }{\max_{i=1,...,N}{|f_i|}},
\label{eq:bayarea_l2err}
\end{equation} 
where $f$ denotes a hydrodynamic field obtained in the reference simulation that resumed from the original restart data,
and $\check{f}(\varepsilon)$ stands for the respective field computed starting from the compressed data.
To simplify the notation, $f$ and $\check{f}(\varepsilon)$ are treated as one-dimensional arrays.

The relative $L^\infty$ error is of order 100\% in both cases for most of the
field variables, except for pressure fluctuation that reaches 14\% and for the base density and pressure, both of which are constant in time and therefore remain of the same order as $\varepsilon$ or less.
The relative $L^2$ error is, in general, two orders of magnitude smaller than the respective $L^\infty$ error,
which means that, in most part of the domain, the pointwise residual error is much smaller than the respective peak values.

Comparing the present results with the global numerical simulation described in Section~\ref{sec:tymip_restart},
one of the key differences is in the spatial resolution. In this building-resolving simulation, the eddy turnover time of 
the smallest wake vortices is of order of seconds. Therefore, after 10 minutes (i.e., by the end of the simulation),
the small structures de-correlate, producing large pointwise error.

\begin{table}[!ht]
\centering
\caption{Relative error after restart. fl: longitudinal momentum; fp: latitudinal momentum; fr: altitudinal momentum; ro: density fluctuation;
ps: pressure fluctuation; rqv: water vapor density; rqq: subgrid scale turbulence kinetic energy; 
surf: surface flux variable.}
\begin{tabular}{ l l l l l }
\hline
Field & $e_\infty(\varepsilon=10^{-6})$ & $e_2(\varepsilon=10^{-6})$ & $e_\infty(\varepsilon=10^{-12})$ & $e_2(\varepsilon=10^{-12})$ \\ \hline
fl    & $1.1901$                & $2.1675 \times 10^{-2}$ & $0.7986$                 & $5.0943 \times 10^{-2}$  \\
fp    & $1.0919$                & $2.6380 \times 10^{-2}$ & $1.0227$                 & $6.6958 \times 10^{-2}$  \\
fr    & $1.0423$                & $2.4495 \times 10^{-2}$ & $1.3071$                 & $7.1998 \times 10^{-2}$  \\
ro    & $0.2440$                & $1.6876 \times 10^{-3}$ & $0.5434$                 & $3.1061 \times 10^{-3}$  \\
ps    & $0.1189$                & $9.8013 \times 10^{-4}$ & $0.1432$                 & $6.0759 \times 10^{-3}$  \\
rqv   & $0.0849$                & $2.3355 \times 10^{-4}$ & $0.5417$                 & $8.5168 \times 10^{-4}$  \\
rqq   & $0.4826$                & $8.8831 \times 10^{-4}$ & $0.6498$                 & $1.4386 \times 10^{-3}$  \\
surf  & $0.6702$                & $2.8779 \times 10^{-3}$ & $0.7086$                 & $5.1263 \times 10^{-3}$  \\ \hline
\end{tabular}
\label{table_bay_err}
\end{table}

\begin{figure}
    \begin{center}
        \includegraphics[scale=0.7]{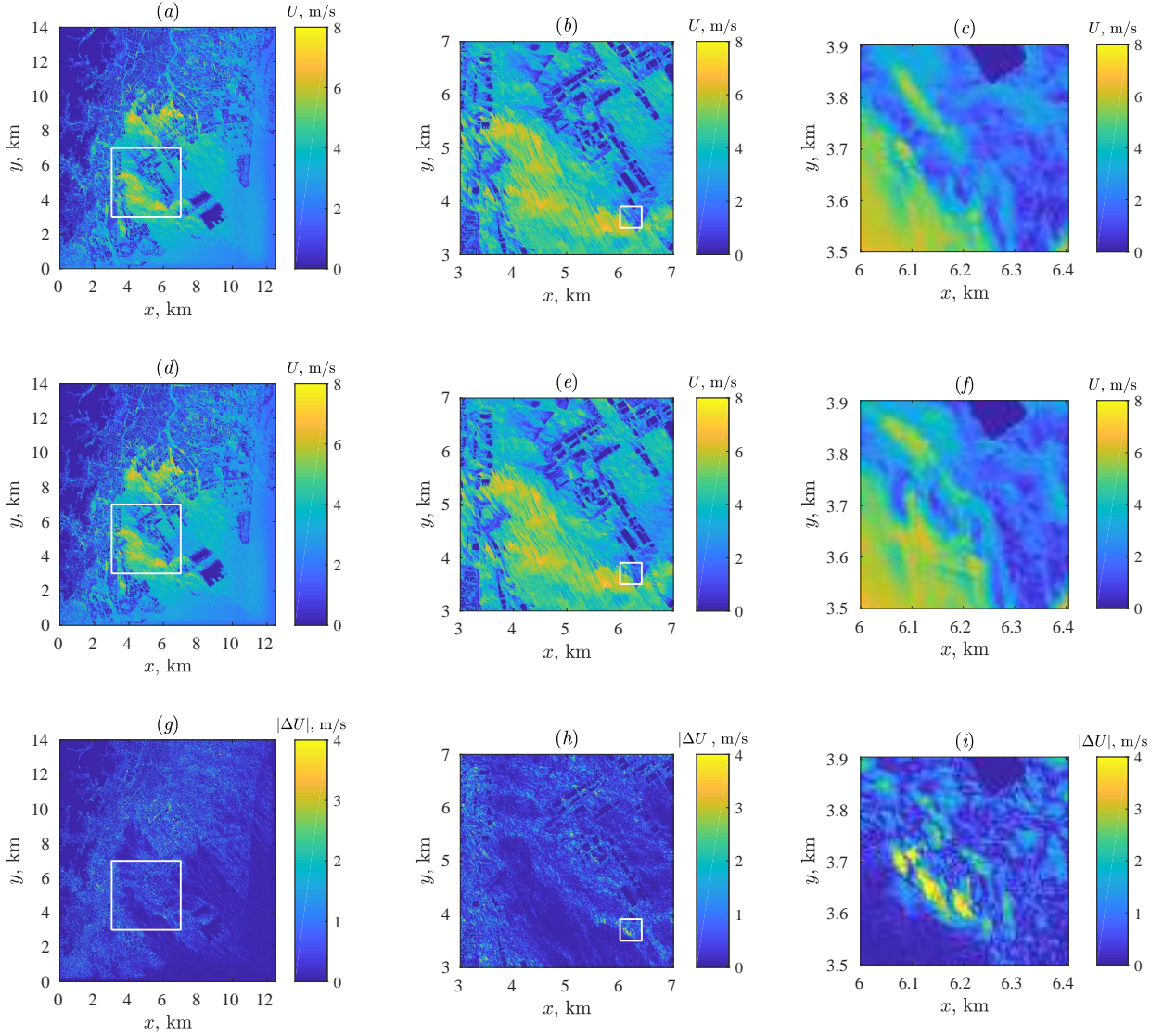}                      
    \end{center}
    \caption{Wind speed distribution in the urban weather simulation, on a plane at 20~m altitude above the sea level.
    Left column shows the entire horizontal span of the computational domain, middle column shows a zoom
    on a selected 4~km $\times$ 4~km area, right column shows a deeper zoom on a 400~m $\times$ 400~m area.
    (\textit{a}), (\textit{b}), (\textit{c}) Restart from the original initial data, results shown for 16:10 JST.
    (\textit{d}), (\textit{e}), (\textit{f}) Restart from the compressed input data using $\varepsilon=10^{-6}$, results shown for 16:10 JST.
    (\textit{g}), (\textit{h}), (\textit{i}) Difference between the results using the original and the compressed initial data.
    Note that the color scale of the bottom-row figures has been adjusted to accentuate on the regions of large $|\Delta U|$.}
    \label{fig:visu_urb}
\end{figure}

\begin{figure}
    \begin{center}
        \includegraphics[scale=0.7]{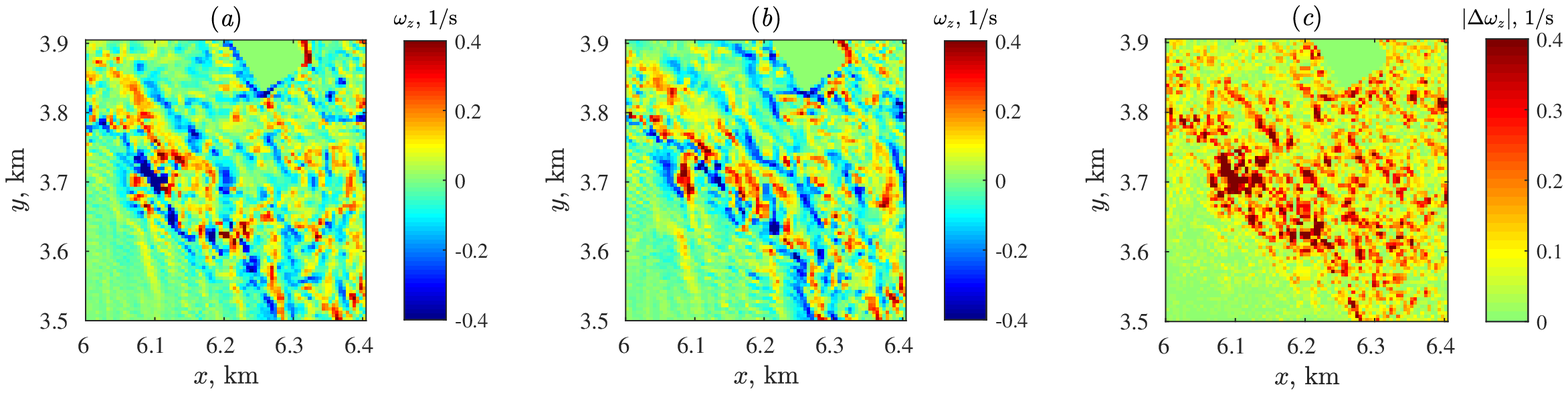}                      
    \end{center}
    \caption{Vertical vorticity component at 20~m altitude above the sea level,
    in a 400~m $\times$ 400~m sub-domain that corresponds to the right column of Fig.~\ref{fig:visu_urb}. All data are for 16:10 JST.
    (\textit{a}) Restart from the original initial data, 
    (\textit{b}) restart from the compressed input data using $\varepsilon=10^{-6}$,
    (\textit{c}) difference between the results using the original and the compressed initial data.}
    \label{fig:visu_urb_vorzoom}
\end{figure}

To gain better insight, let us consider a horizontal plane at 20~m altitude above the sea level.
Figure~\ref{fig:visu_urb} shows the 
velocity magnitude distribution $U$ in different cases.
The top row panels (\textit{a}), (\textit{b}) and (\textit{c}) correspond to the result at 16:10 JST 
of the simulation resumed from the original restart data.
Panel (\textit{a}) displays the entire slice while (\textit{b}) and (\textit{c}) zoom on selected sub-domains.
The result of a restart with $\varepsilon=10^{-6}$ is shown in Figs.~\ref{fig:visu_urb}(\textit{d}), (\textit{e}) and (\textit{f}).
Large-scale structures are essentially the same as in Figs.~\ref{fig:visu_urb}(\textit{a}) and (\textit{b}). However, a careful examination
reveals significant differences on a smaller scale, compare between panels (\textit{f}) and (\textit{c}). 
To focus on such small-scale discrepancy, we calculate 
the difference between the velocity fields obtained with and without compression, $|\Delta U| = |U(\varepsilon=10^{-6}) - U(\varepsilon=0)|$, and display it in 
Fig.~\ref{fig:visu_urb}(\textit{g}), (\textit{h}) and (\textit{i}) on an exaggerated color scale.
The darker tone of panels (\textit{g}) and (\textit{h}) suggests that $|\Delta U|$ is generally much smaller than $U$.
There are, however, many bright spots around the buildings that mark the small-scale differences in the wake.
A zoom on one of these spots displayed in Fig.~\ref{fig:visu_urb}(\textit{i}) reveals that, 
locally, $|\Delta U|$ is of the same order of magnitude as $U$,
in agreement with the global $L^\infty$ error evaluations shown in table~\ref{table_bay_err}.
In addition, the error is spatially organized in a pattern characteristic of mixing layers.
The vorticity plots in Fig.~\ref{fig:visu_urb_vorzoom}(\textit{a}), (\textit{b}) and (\textit{c})
show that $\omega_z$ is small in the bulk of the fast current (lower-bottom corner of the subdomain), 
but many strong small-scale vortices are present in the mixing layer as well as in the slow current around the buildings.
Although these small vortices show qualitatively similar arrangement in Fig.~\ref{fig:visu_urb_vorzoom}(\textit{a}) as in Fig.~\ref{fig:visu_urb_vorzoom}(\textit{b}),
the exact position differs by as much as the core size. Consequently, the error $|\Delta \omega_z| = |\omega_z(\varepsilon=10^{-6}) - \omega_z(\varepsilon=0)|$
is a superposition of strong well-localized peaks.
It is worth mentioning that a restart with $\varepsilon=10^{-12}$ has led to very similar results. 
This is expected as small-scale vortices shed from the buildings evolve rapidly and chaotically.
Exact deterministic repetition of such simulation requires that the initial data be exact.
On the other hand, the initial error has negligible effect on a kilometer scale.

\section{\label{sec:cost}Compression performance with consideration of computational cost}

Our intention is to compress data from numerical simulations. 
Let us first consider a scenario when the same computer is used for the simulation and for the data compression.
Writing the output data in a divided file, as explained in Section~\ref{sec:bayarea}, enables parallel execution of the compression program, 
albeit at a cost of slight increase in the compressed file size.
Simulations often use time-marching schemes, each time iteration incorporates differential operators
of linear computational complexity (i.e., proportional to the number of grid points $N$) in the case of, e.g., evaluating derivatives using finite differences, or having an even higher complexity (e.g., $N \log{N}$ or $N^2$) if spectral methods are used or linear systems are to be solved.
Although wavelet transform and range coding are known to be relatively expensive operations, their computational complexity is linear.
This means that the number of arithmetic operations necessary to compress one snapshot of the output data is, at worst, proportional to the number of arithmetic operations required for one time iteration in the simulation. 
The proportionality constant may be greater than one if the simulation only uses explicit low order schemes and the right-hand side of the evolution equation is simple enough. However, typical practical problems in scientific computing are computationally more intense than compression using one pass of wavelet transform followed by quantization and range coding.
Besides that, it is rare to write on disk the output after every time step. Temporal sampling is commonly used \cite{Li_2018_cgf}, which dramatically reduces 
the data compression cost in comparison with the simulation cost. 

A different scenario would be to perform a simulation on a supercomputer and compress/decompress the result files on a desktop computer.
The elapsed time of data compression can be long in such situation, and that is the case we focus on in this section.
For the performance analysis, we use the HIT dataset described in Section~\ref{sec:hit_compression}.
The simulation was performed on the Earth Simulator supercomputer system (NEC SX-ACE),
while the compression/decompression analyzed on an HP Z640 workstation with two Intel Xeon E5-2620v4 8-core CPUs at $2.10$GHz clock rate, $8 \times 8$GB DDR4-2400 RAM and a RAID 1 pair of Seagate $2$TB 7200 RPM SATA HDDs.
A succinct discussion of the WaveRange software performance and optimization aspects
will be followed by a comparison with other open-source libraries.

\subsection{\label{sec:roofline}Performance assessment using a roof-line model}

\begin{figure}
    \begin{center}
        \includegraphics[scale=0.7]{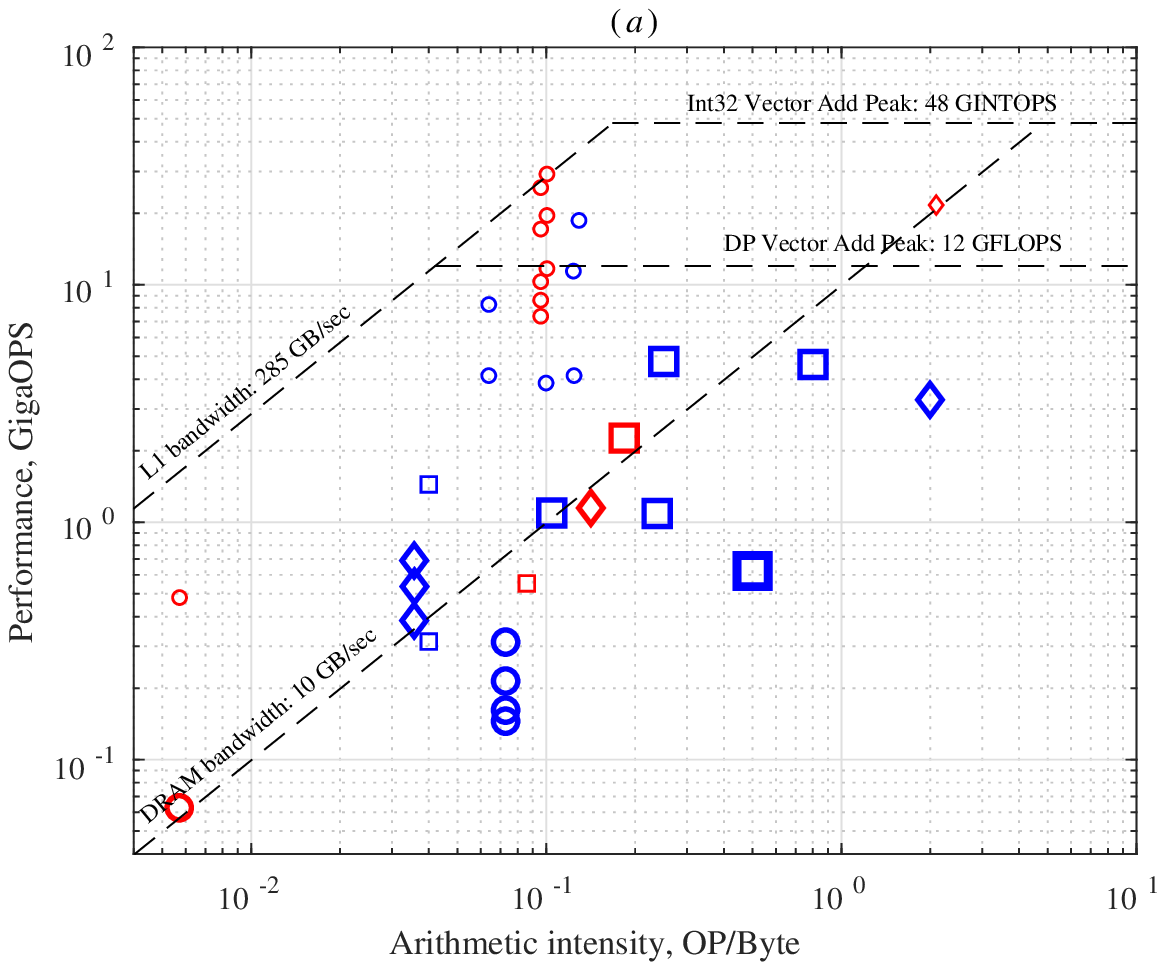}\includegraphics[scale=0.7]{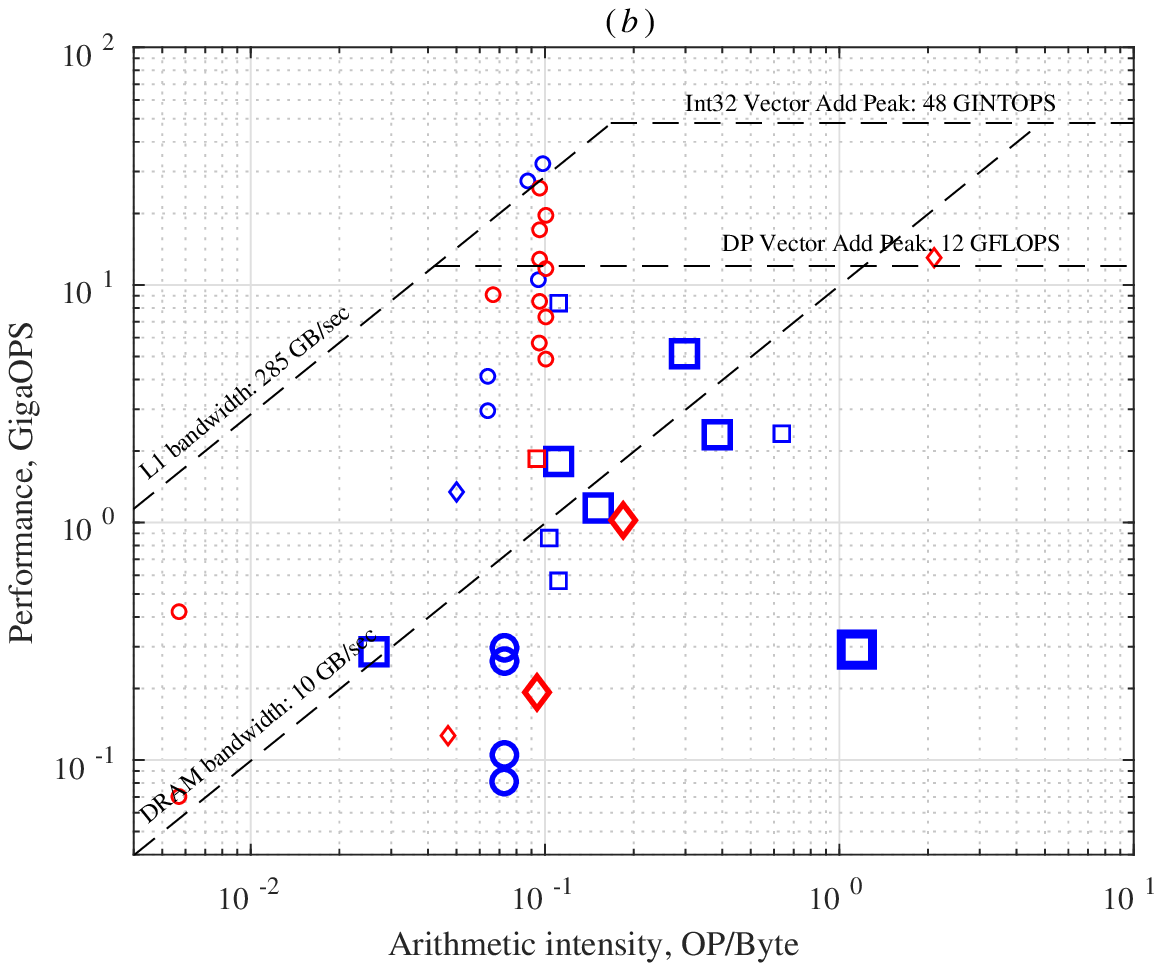}
    \end{center}
    \caption{Roofline model.
    (\textit{a}) Compression performance;
    (\textit{b}) Decompression performance.
    Circles show loops that belong to the wavelet transform routine, squares correspond to range encoding/decoding and diamonds belong to 
    quantization/dequantization or the rest of the program. Blue markers show scalar loops and red markers show vectorized loops. 
    The size of the marker and its line width signifies its relative contribution to the program total elapsed time: less than 1\% for the smallest markers, 
    between 1\% and 15\% for the medium-size markers and more than 15\% for the large markers. Markers may overlap.
    }
    \label{fig:roofline}
\end{figure}

We used Intel Advisor 2019 Update 4 (build 594873) for the performance analysis.
WaveRange was built using gcc 7.4.0 with -O2 level optimization and AVX2 support, under the Ubuntu 18.04.1 operating system.
The $x$-velocity component of the HIT dataset was read from a file in the FluSI HDF5 format, compressed with $\epsilon=10^{-6}$ and written on disk also using HDF5.
Then it was decompressed.
The compression and the decompression programs have been profiled separately. Therefore, 
the results are presented in two columns in table~\ref{table_perf} and in two panels in figure~\ref{fig:roofline}.

\begin{table}
\centering
\caption{Performance characteristics of WaveRange as applied to the HIT dataset.}
\begin{tabular}{ l l l }
\hline
                                       & Compression & Decompression \\ \hline
Total CPU time                         & $20.63$~s   & $24.06$  \\
\quad\emph{Wavelet transform}          & $23$\%      & $26$\%  \\
\quad\emph{Range coding}               & $41$\%      & $61$\%  \\
\quad\emph{Other (incl. qunatization)} & $36$\%      & $13$\%  \\
Time in vectorized loops               & $11$\%      & $12$\%  \\
Time in scalar code                    & $89$\%      & $88$\%  \\
\end{tabular}
\label{table_perf}
\end{table}

In the compression procedure, the wavelet transform is the least time consuming step.
This can be explained by the relatively large fraction of vector operations in it: there are 12 vector loops 
and 12 scalar kernels shown in figure~\ref{fig:roofline} with the red and the blue circles, respectively.
Two of them achieve the L1 bandwidth bound.
The need for strided memory access is the main factor that limits further optimization of the transform (these loops are marked with blue circles 
situated below the DRAM bandwidth line).
In contrast, only 1 of the 8 range coder kernels has been vectorized, and the most time-consuming one is the function
that encodes a symbol using frequencies. This explains why range coding is the most time consuming step.
Quantization and other parts of the program count 2 vector loops and 4 scalar kernels.
In the quantization process, type conversion from double to char presents difficulties for the automatic optimization.
The overall time in vectorized loops amounts to 11\% of the total CPU time.

Decompression with WaveRange takes longer time than compression.
Although the inverse wavelet transform has 15 vector and 6 scalar kernels, its execution takes longer than the forward transform used in the compression.
Range decoding is by far the most expensive part, as it amounts to 61\% of the decompression CPU time. 
Only 1 of the 10 range coder kernels has been vectorized.
The most time consuming function is the one that calculates cumulative frequency of the next symbol. 
Dequantization is relatively cheap: it only takes 13\% of the total CPU time.
The overall vector time ratio is 12\%, which is nearly the same as in the compression program.

\subsection{\label{sec:comparison}Comparison with other methods}

Before comparing WaveRange with other methods, it is important to provide baselines in terms of lossless compression achievable
with general-purpose utilities. This is summarized in table~\ref{table_lossless}.
As in the previous section, here we use the $x$-velocity component of the HIT dataset.

DEFLATE \cite{Deutsch_1996_techrep} is an algorithm widely used for general purposes. The best compression, i.e., the smallest compressed file size, 
is achieved when the level of compression is set to 9. The execution time can be minimized by setting the level of compression to 1.
It is customary to apply shuffling as a pre-conditioner to facilitate floating point data compression with DEFLATE.
The idea of shuffling is to break apart floating point elements of an array into their mantissa and exponent components, then
change the byte order in the data stream such as to   
place the first byte of every element in the first chunk, then the second byte of every element in the second chunk, etc.
In the scientific datasets, values at neighboring points are usually close to each other.
Therefore, shuffling produces a data stream that includes many continuous sub-sequences of identical entries, which DEFLATE can compress well.
From this point of view, shuffling is similar to the quantization described in Section~\ref{sec:quantization}.

The last two lines in table~\ref{table_lossless} show the performance of Szip and 7-Zip.
Szip \cite{Yeh_2002_estc} is an implementation of the extended-Rice lossless compression algorithm designed for use with scientific data. In our test, we activated the optional nearest neighbor coding method.
7-Zip applies the LZMA method \cite{Pavlov_2019_techrep}, and we set the level of compression to 9, which is the maximum level. 
The crosses in figure~\ref{fig:hit_res}(\textit{b}) correspond to the compressed file size obtained with this method.
All algorithms have been applied as HDF5 filters (HDF5 version 1.10.0-patch1), with the exception of 7-Zip, which was used as a standalone application (version 16.02).

\begin{table}
\centering
\caption{Lossless compression of the HIT data: relative compressed file size $\Sigma$, compression ratio $r$, compression throughput $\theta_c$ and decompression throughput $\theta_d$}
\begin{tabular}{ l l l l l }
\hline
                               & $\Sigma$  & $r$     & $\theta_c$  & $\theta_d$ \\ \hline
DEFLATE, fastest               & $96.3$\% & $1.038$ & $23.2$~MB/s & $144.1$~MB/s  \\
DEFLATE, best                  & $95.6$\% & $1.046$ & $16.9$~MB/s & $153.0$~MB/s  \\
Shuffling and DEFLATE, fastest & $82.6$\% & $1.211$ & $31.0$~MB/s & $282.6$~MB/s  \\
Shuffling and DEFLATE, best    & $81.2$\% & $1.231$ & $12.7$~MB/s & $290.2$~MB/s  \\
Szip                           & $84.4$\% & $1.184$ & $61.3$~MB/s & $124.8$~MB/s  \\
7-Zip                          & $87.4$\% & $1.144$ & $1.6$~MB/s  & $17.0$~MB/s  \\
\end{tabular}
\label{table_lossless}
\end{table}

The performance metrics are the relative compressed file size $\Sigma$ as defined by (\ref{eq:sigma}), compression ratio $r$ as defined by (\ref{eq:crt}), compression throughput $\theta_c = s(f)/t_c$ and decompression throughput $\theta_d = s(f)/t_d$, where $s(f)=1$~GB is the storage space required for a $512^3$ array
of double-precision values, $t_c$ is the CPU time for compression and $t_d$ is the CPU time for decompression.
DEFLATE with the lowest level of compression reduced the file size by only a very small amount, down to $96.3$\% of the original size.
Switching to the maximum level of compression helps to gain additional $0.7$\%, but shuffling the data brings a dramatic improvement, 
allowing to reach $81.2$\%, which is the best result among all lossless methods considered here.
Szip is slightly less efficient from the file size reduction viewpoint, but its compression throughput is higher. However, the decompression throughput is lower for 
Szip than for DEFLATE. 7-Zip turns out to be less efficient by all metrics, which is not surprising, since it is not
designed for use with scientific data.
Overall, the compression ratio is $1.231$ at most, meaning that many-fold data reduction cannot be achieved without loss of information in this example.

\begin{figure}
    \begin{center}
        \includegraphics[scale=0.7]{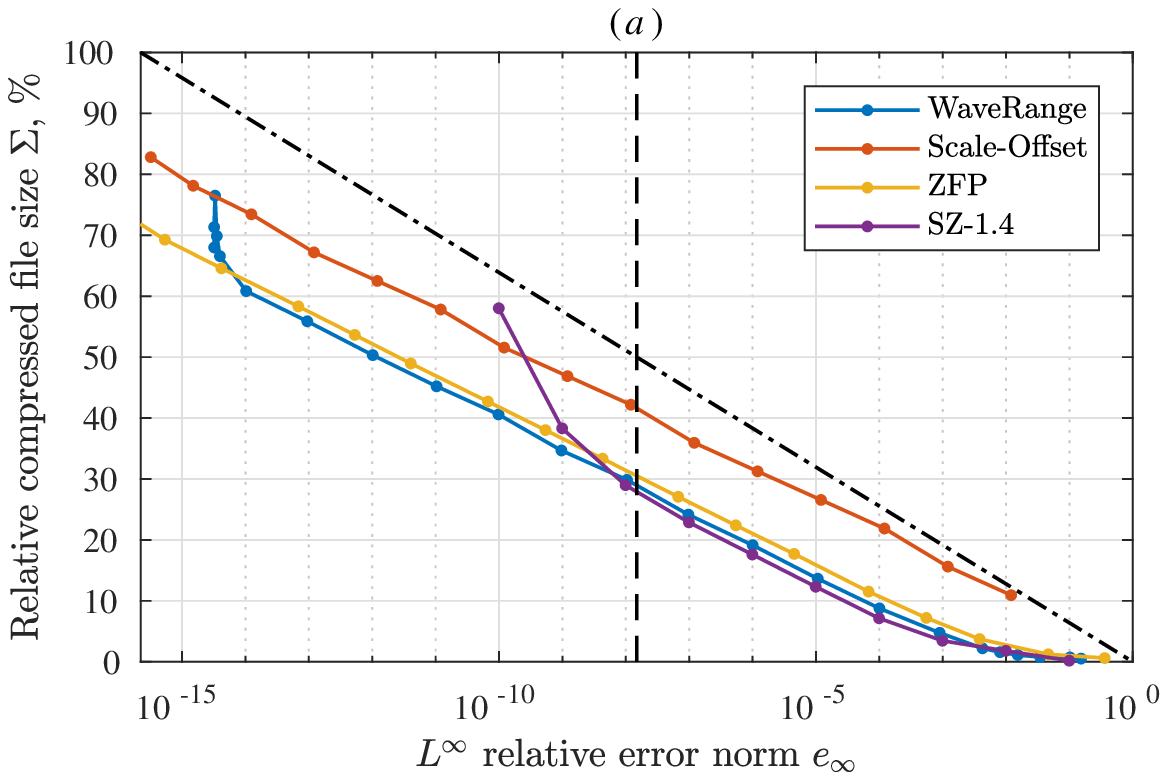}\includegraphics[scale=0.7]{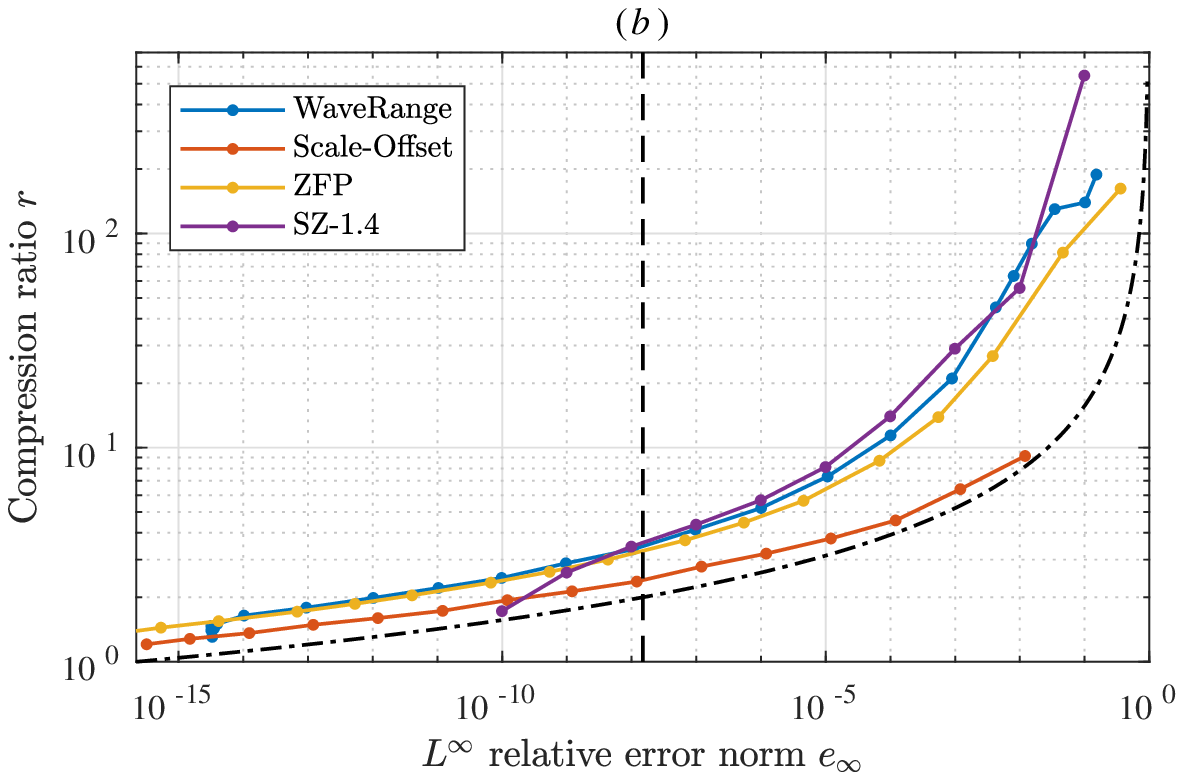}
        \includegraphics[scale=0.7]{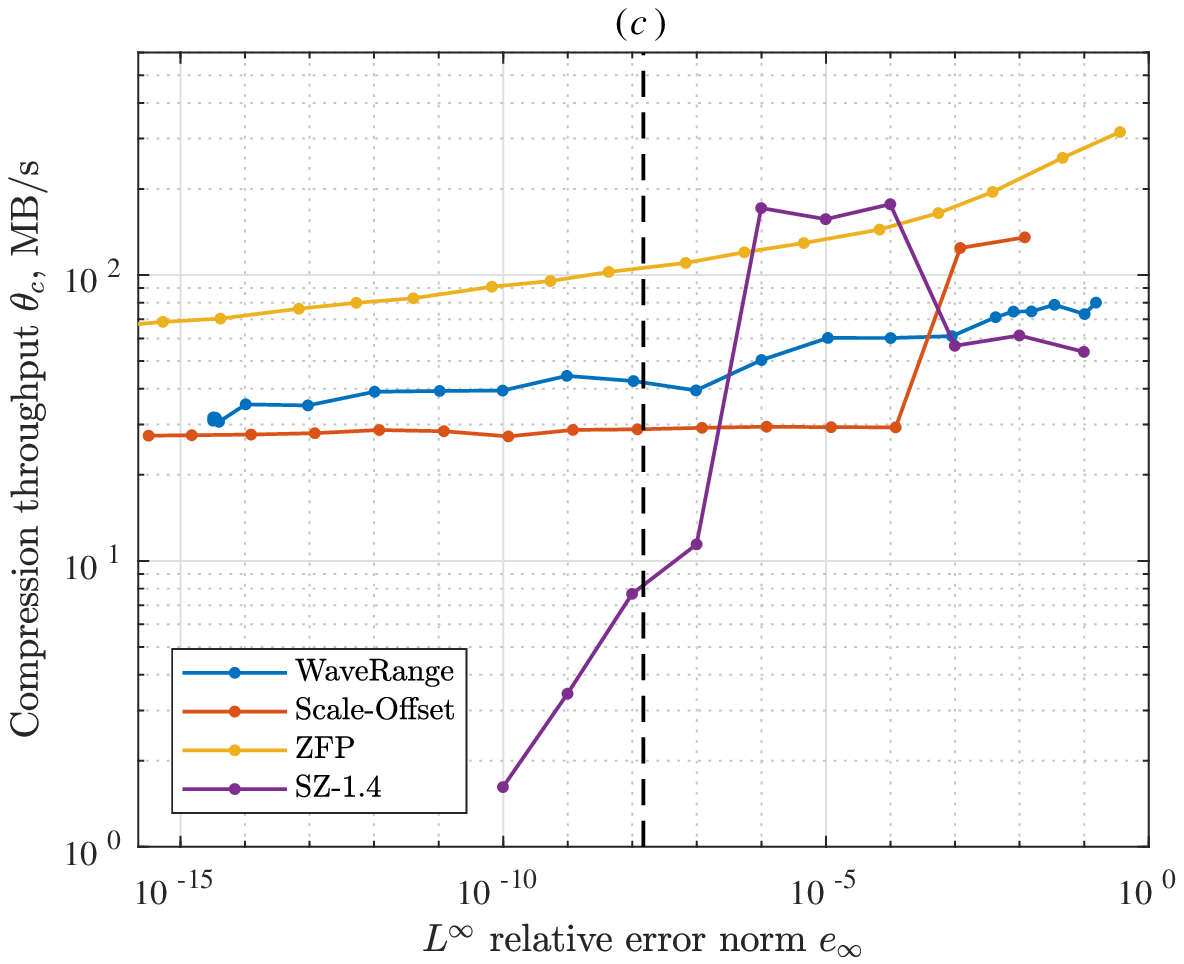}\includegraphics[scale=0.7]{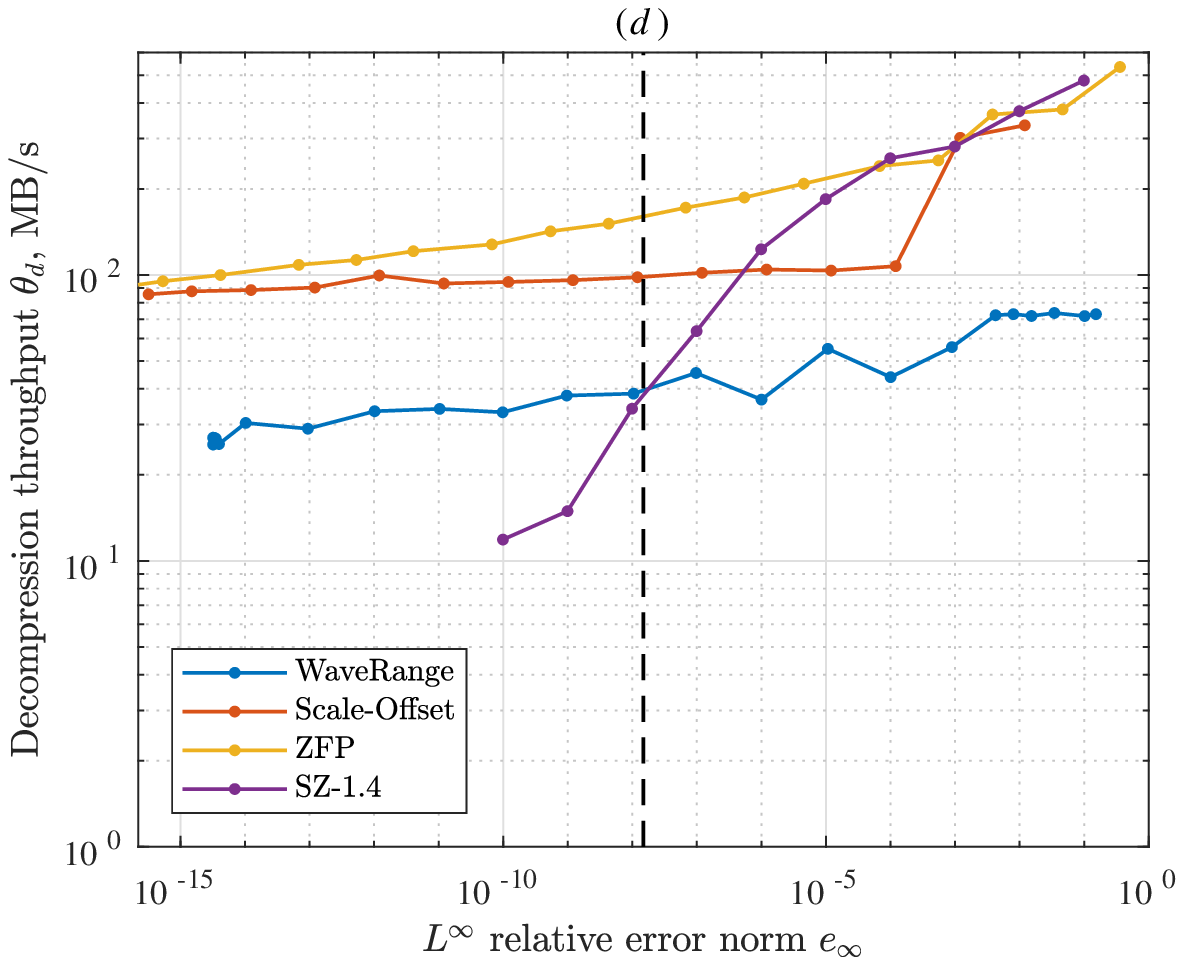}        
    \end{center}
    \caption{Comparative performance of different scientific data compression methods.
    (\textit{a}) Compressed file size in per cent of the original file size;
    (\textit{b}) Compression ratio;
    (\textit{c}) Compression throughput;
    (\textit{d}) Decompression throughput.
    The dashed vertical line shows the accuracy of single-precision storage and
    the dash-dot lines correspond to compression by quantization only.}
    \label{fig:bench}
\end{figure}

The performance of different lossy compression methods, including WaveRange, is displayed in figure~\ref{fig:bench},
which shows $\Sigma$, $r$, $\theta_c$ and $\theta_d$ as functions of the $L^\infty$ error norm.
Three other lossy methods have been evaluated. Scale-Offset is an HDF5 filter that performs a scale and offset operation, then truncates the result to a minimum number of bits.
Scaling is performed by multiplication of each data value by 10 to the power of a given scale factor,
which can be adjusted in order to reach the desired compressed file size or accuracy.
SZ-1.4 \cite{Tao_2017_ieee} and ZFP \cite{Diffenderfer_2019_sisc} are two state-of-the-art scientific data compressors.
The former is based on the Lorenzo predictor and the latter uses a custom transform that operates multi-dimensional blocks of small size.
Both are implemented as HDF5 filters and both can operate in a fixed error mode.
To obtain the plots in figure~\ref{fig:bench}, we followed the same procedure as described in the previous sections:
the HIT $x$-velocity data file was compressed, the file size was measured, then it was decompressed and the $L^\infty$ error was measured.
In addition, the execution time was measured for the compression and for the decompression separately.

Let us first discuss the compression performance in terms of $\Sigma$ and $r$.
The scale-offset compression generally produces larger files than WaveRange.
Note that this method is equivalent to quantization (\ref{eq:quant}) in the limit of large $L^\infty$ error.
ZPF produces slightly larger files than WaveRange, except when the set tolerance is smaller than $10^{-14}$
and the $L^\infty$ error of WaveRange increases significantly due to roundoff.
The performance of SZ-1.4 varies largely depending on the control parameter configuration.
In our test, we optimized it for the maximum compression ratio at a given error magnitude.
Thus, the maximum quantization interval number was increased as the error decreased.
With this setting, SZ-1.4 could produce smaller files than WaveRange, but only for the relative error greater than $10^{-8}$.
SZ-1.4 switched to lossless mode when the relative error smaller than $10^{-10}$ was requested.

Considering the throughput $\theta_c$ and $\theta_d$, ZFP was generally the fastest in our tests,
although it was outperformed by SZ-1.4 in certain cases by a small amount.
WaveRange was 3 times slower than ZFP for the compression and 5 times slower for the decompression.
This is not surprising, considering that the transform used in ZFP was optimized for the maximum throughput 
at the cost of a certain decrease in the compression ratio.
Interestingly, despite the relatively high algorithmic complexity, WaveRange compression was faster then the scale-offset compression.

\section{\label{sec:conclu}Conclusions}

A wavelet-based method for compression of 
data output from numerical simulation of fluid flows using block Cartesian grids has been presented. 
The method consists of a discrete wavelet transform, quantization adapted for floating-point data, and entropy coding. 
It is designed such as to guarantee the desired pointwise reconstruction accuracy.
An open-source software implementation has been provided, see \url{https://github.com/pseudospectators/WaveRange}.

The data compression properties have been analyzed using
example numerical tests from different kinds of problems, from idealized fluid flows to realistic seismology and weather simulations.
In particular, it is found that, in the most challenging (from the compression point of view) case of homogeneous isotropic turbulence,
compression allows to reduce the data storage by a factor of 3 using $\varepsilon=10^{-8}$, which is significantly better than 
storing the same data in single-precision floating-point format.
The method show favorable scaling with the data size, i.e., greater compression ratios are achieved for larger datasets.
Compression of the wake turbulence is also slightly better, compared to the reference homogeneous isotropic turbulence case.
This is explained by greater inhomogeneity of the wake turbulence, which means that there are fewer large wavelet coefficients.

The compression performance depends on the flow type. For the realistic data generated from global and urban weather simulations, the file size can be reduced by a factor of about 15 using
the threshold value $\varepsilon=10^{-6}$.
Both simulations can be successfully restarted from the reconstructed data.
The reconstruction error has shown no significant effect on the dynamics of large-scale structures,
which are typically the main objects of interest.
It should be noted, however, that the small-scale structures may randomize very quickly if
any small initial error is introduced in the simulation.

\section*{\label{sec:data}Data availability}

The HIT and the wake turbulence datasets, in the compressed format, are available at \url{https://osf.io/pz4n8/}.
In addition, the same HIT dataset is contained in the supplementary file \verb|data_sample.zip|.
Access to the weather simulation data can be granted upon request, under a collaborative framework between JAMSTEC and related institutes or universities.

\section*{Acknowledgements}
\label{}

This work is supported by the FLAGSHIP2020, MEXT within the Post-K Priority Issue 4 (Advancement of meteorological and global environmental predictions utilizing observational ``Big Data"). The authors thank Dr. Koji Goto and Dr. Keigo Matsuda for their help with handling the global- and urban-scale simulations,
and Prof. Seiji Tsuboi for providing the seismology simulation data.


\appendix
\section{\label{sec:wake_compression}Wake turbulence}

The fluid velocity field in this case is obtained from a numerical simulation of 
viscous incompressible flow past a periodic array of circular cylinders.
The flow configuration is similar to the numerical wind tunnel with a cylindrical obstacle considered in \cite{Ravi_2016_srep}. Here it is described in dimensionless units.
The fluid domain is a rectangular box with length of 10, width of 8 and height of 4.
The boundary conditions on the exterior faces of the domain are periodic in all the three directions. In addition to that, 
a vorticity sponge condition is applied over the 48 grid slabs adjacent to the outflow boundary,
to avoid the wake re-entering the domain. 
The cylinder immersed in the fluid has the diameter of 1.894. Its axis is oriented vertically and it is located at 1.5 length units downstream from the upstream boundary
of the periodic domain. The fluid has the kinematic viscosity of 0.001 and density of 1.
Mean inflow velocity of 1.246 is imposed. 

The computational domain is discretized using a uniform Cartesian grid consisting of $960 \times 768 \times 384$ points.
The no-slip boundary condition at the surface of the cylinder is modeled using the volume penalization method
with the penalization parameter $C_\eta=5 \cdot 10^{-4}$.  
For more information about the numerical method, see \cite{Engels_2016_sisc}.

A small cylindrical detail attached on the surface of the cylinder at the angular distance of $135^\circ$ from the front stagnation point served to quickly break the bilateral symmetry of the flow.
Subsequently, random noise introduced during the startup phase provoked three-dimensional instabilities.
The three components of the velocity field ($u$,$v$,$w$) at time $t=330$ were saved, respectively, in three separate files
in double precision. Each file occupied 2.2~GB of hard disk space.

\begin{figure}
    \begin{center}
        \includegraphics[width=0.55\textwidth,clip]{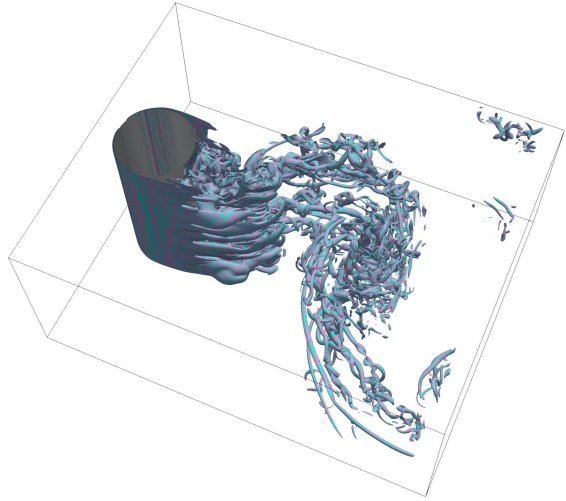}
    \end{center}
    \caption{Flow visualization of the turbulent wake simulation. 
    Black lines show the extent of the computational domain.
    The vorticity magnitude is visualized using an iso-surface at 10\% of the maximum value.
	Iso-surfaces of the
    original field (cyan) and reconstructed after compression with $\varepsilon=10^{-2}$ (magenta) 
    are superimposed and overlap almost perfectly. The surface of the cylinder is colored in grey.}
    \label{fig:cyl}
    \vspace*{-3mm}
\end{figure}

The flow configuration and the result of the simulation are visualized in Fig.~\ref{fig:cyl}.
The wake is apparently turbulent with a variety of scales and heterogeneity reminiscent of industrial flows.
Grey color shows the cylinder, cyan shows an iso-surface of the vorticity magnitude
calculated using the original velocity field, and magenta shows a similar iso-surface for $\varepsilon=10^{-2}$.
The two iso-surfaces overlap.

\begin{figure}
    \begin{center}
        \includegraphics[scale=0.7]{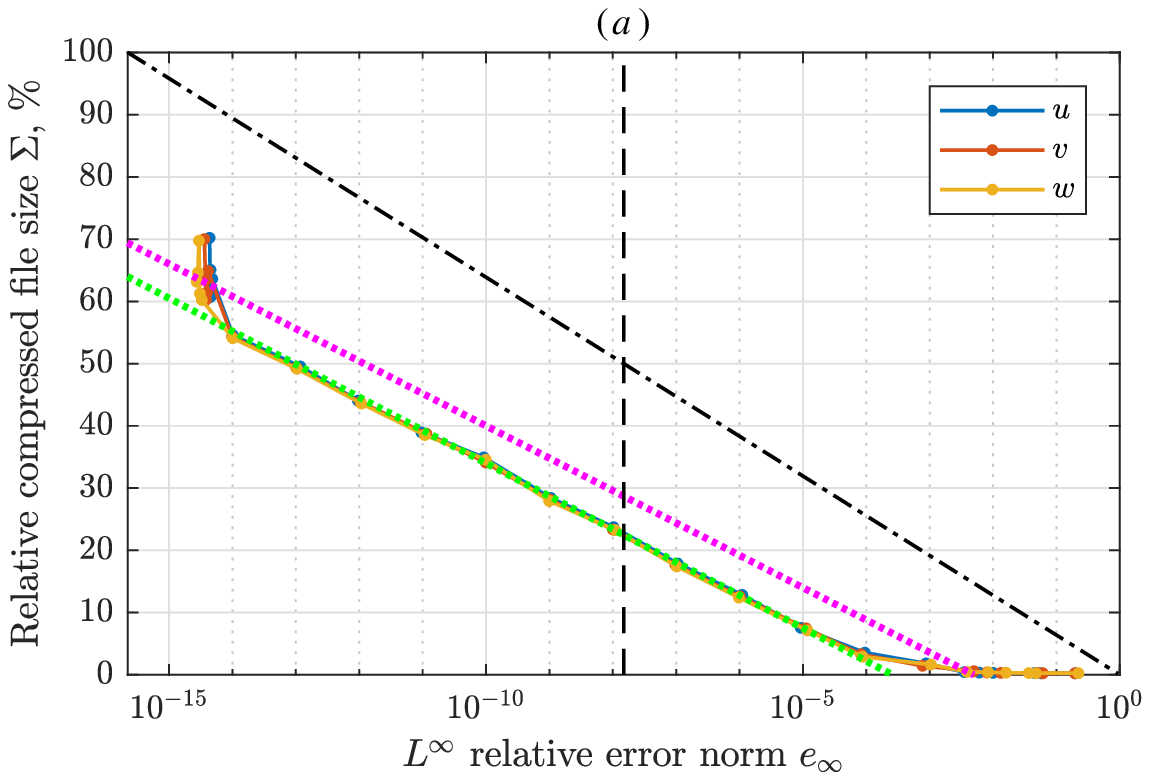}
        \includegraphics[scale=0.7]{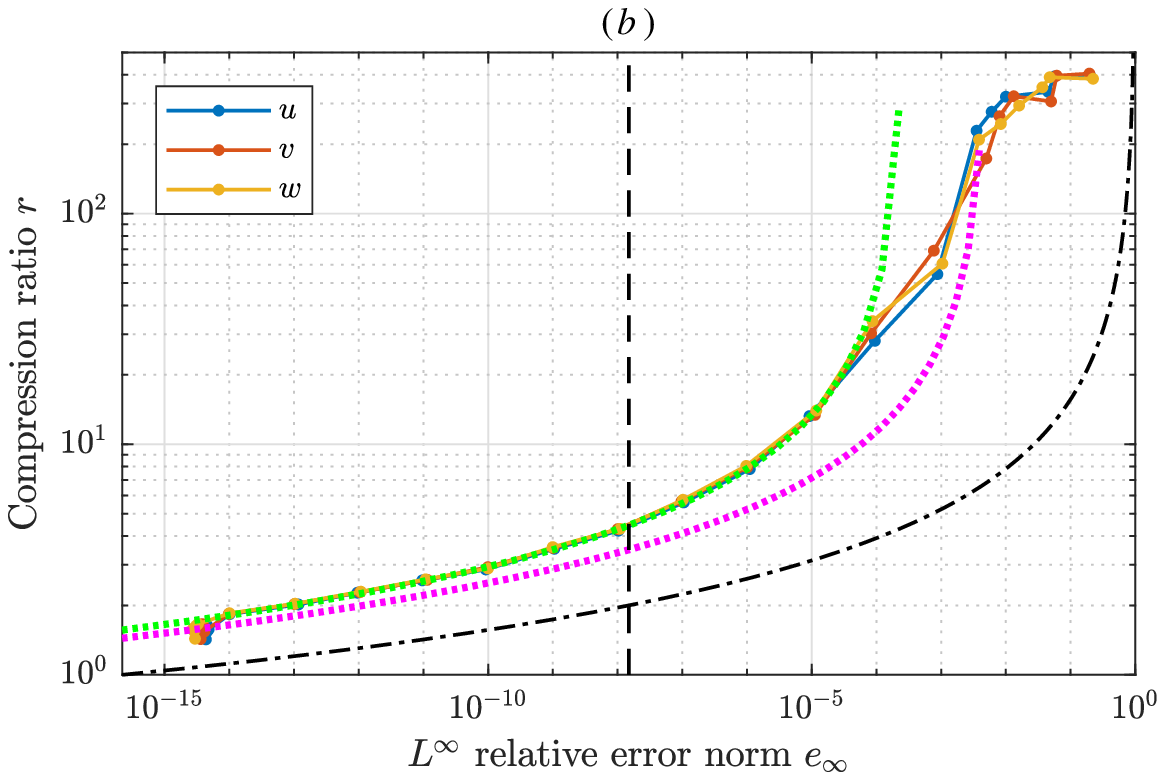}                
    \end{center}
    \caption{Compression of the wake velocity components.
    (\textit{a}) Compressed file size in per cent of the original file size, as a function of the $L^\infty$ error norm;
    (\textit{b}) Compression ratio versus the relative $L^\infty$ error.
    The fits (\ref{eq:sigma_hit}) and (\ref{eq:sigma_cyl}) that correspond to $\Sigma_{HIT}$ and $\Sigma_{Cyl}$, respectively, 
    are shown with the magenta and green dotted lines.
    The dashed vertical line shows the accuracy of single-precision storage and
    the dash-dot lines correspond to compression using quantization only.}
    \label{fig:wake}
\end{figure}

Similarly to the previous HIT test case, the procedure of compression with prescribed tolerance $\varepsilon$
and subsequent reconstruction has been applied to the velocity components
of the turbulent wake. Again, 
the relative error measured in $L^\infty$ norm
has appeared almost identical to $\varepsilon$, except for the smallest and for the largest $\varepsilon$.
Fig.~\ref{fig:wake}(\textit{a}) shows the compressed file size as a function of the error norm, component-wise.
The compressed file size is again normalized with the original file size. The error norm is defined by (\ref{eq:errlinf}). The compression method is equally effective for all velocity components, despite the
anisotropy of the velocity field.
Perfect reconstruction cannot be reached because of round-off errors, but reconstruction with $10^{-14}$ accuracy is achieved 
from a compressed file slightly larger than one half of the original size. It can also be noticed that,
if the velocity field were stored in a single precision file, the error would be 100,000 times larger than when storing the same field
using the compressed format in an equally large file. 
Similarly to Fig.~\ref{fig:hit_res}(\textit{b}) for the HIT test case, the diagonal dash-dotted line in Fig.~\ref{fig:wake}(\textit{a}) shows the gain in compression achieved by discarding the least significant digits, i.e., 
by reducing the precision of each point value. The difference between this upper bound and the actually measured file size is the combined effect of wavelet transform and entropy coding. 

The compression ratio as a function of the error norm is shown in Fig.~\ref{fig:wake}(\textit{b}).
Greater compression ratios can be achieved when precision requirements are less stringent. For instance, in this example
one can achieve 8 times reduction in the volume of data if stored with $10^{-6}$ accuracy.
For very low accuracy, the compression ratio saturates at $r \approx 400$, which is close to the maximum compression ratio obtained for the HIT data.

\begin{figure}
    \begin{center}
        \includegraphics[scale=0.7]{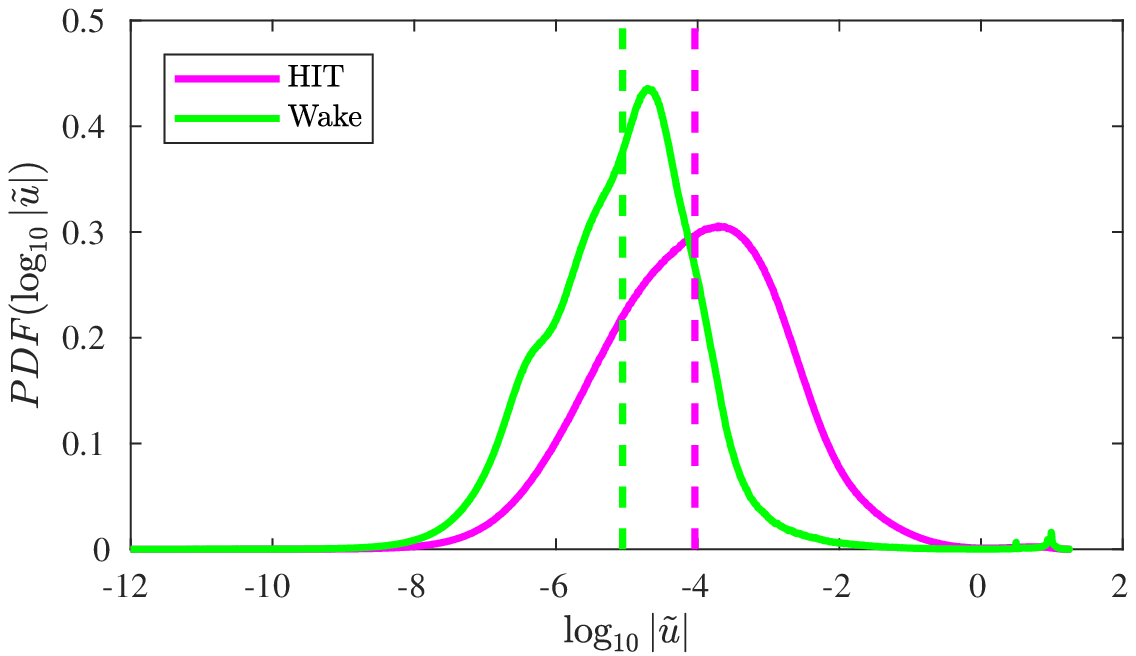}          
    \end{center}
    \caption{Histogram of the order of magnitude of the wavelet coefficients of $u$
    velocity component in two different test cases.}
    \label{fig:pdf}
\end{figure}

In the intermediate range of $\varepsilon$, the compressed file size
as a function of the relative $L^\infty$ error can be approximated as
\begin{equation}
\Sigma_{Cyl} = (-0.19 - 0.053 \log_{10}{e_\infty}) \times 100\%,
\label{eq:sigma_cyl}
\end{equation}
which is shown in Fig.~\ref{fig:wake}(\textit{a}) using green dots.
It can be compared with $\Sigma_{HIT}$ given by (\ref{eq:sigma_hit}), which is superposed on the same figure using magenta dots. The values of $\Sigma_{Cyl}$ are smaller than those of $\Sigma_{HIT}$.
This can be explained by comparing the histograms
displayed in Fig.~\ref{fig:pdf}.
Only the first component, $u$, is shown for clarity. The results for $v$ and $w$ are similar.

To guarantee fair comparison between two histograms for the two different flow fields,
$u$ is normalized by its half-span before applying the wavelet transform yielding $\tilde{u}$.
The same normalization is used in the data compression algorithm.
As the number of bits required to represent a point value of $\tilde{u}$,
after quantization, is proportional to $\log_{10}|\tilde{u}|$,
the latter quantity is used to produce the histogram.
The interval between its minimum and maximum is divided in a finite number of bins and the number of point values falling in each bin is counted. 
Note that the maximum values are almost identical for both datasets.
The result is normalized such that the area under the curve integrates to 1.

By comparing the histograms for the HIT and the wake velocity datasets,
one can see, for example, that 
the HIT field has relatively many coefficients of order of magnitude $10^{-2}$, but less at $10^{-6}$.
The expected value for the HIT case is $-4$, whereas in the cylinder wake case it is equal to $-5.1$.
The standard deviation is similar in both cases: $1.3$ and $1.1$, respectively.
It follows that the HIT wavelet coefficients are, on average, almost one order of magnitude larger
than the cylinder wake wavelet coefficients. Consequently, for equal compression ratio, the 
$L^\infty$ error is expected to be one order of magnitude larger for the HIT data than for the cylinder wake.
This is in agreement with the observed difference between the linear fits in Fig.~\ref{fig:wake}(\textit{a}).
In addition, the slightly larger skewness of the cylinder wake PDF explains why the difference becomes slightly smaller when the tolerance is decreased - also compare the slopes of (\ref{eq:sigma_hit}) and (\ref{eq:sigma_cyl}).

WaveRange treats individual components of a vector field independently.
Although it must be possible to exploit correlation between multiple scalar fields,
this procedure is not straightforward.
We have tested two approaches. The results are compared in table~\ref{table_vector}.
The `Velocity -- polar' method consists in transforming the velocity components $u$, $v$ and $w$ 
to a magnitude $\rho$ and two angles $\vartheta$ and $\phi$ such that
\begin{equation}
u = \rho \sin{\vartheta} \cos{\phi}, \quad v = \rho \sin{\vartheta} \sin{\phi}, \quad w = \rho \cos{\vartheta}.
\end{equation}
The scalar fields $\rho$, $\vartheta$ and $\phi$ are then compressed with a prescribed tolerance $\epsilon$.
The `Vorticity -- Cartesian' method calculates the vorticity $\vec{\Omega} = \vec{\nabla} \times \vec{U}$, 
which is the curl of the velocity vector $\vec{U} = (u,v,w)$, and the spatial average of the velocity $\vec{U}_0$.
Then, the three components of the vorticity are compressed with tolerance $\epsilon$.
The velocity is reconstructed from $\vec{\Omega}$ and $\vec{U}_0$ using the Biot-Savart formula.
All differential and integral operators are approximated using a Fourier spectral method.

Both methods introduce error due to additional computations.
For the `Velocity -- polar' method, computation entails a larger round-off error than in the original `Velocity -- Cartesian' case.
The `Vorticity -- Cartesian' method involves numerical differentiation which has a truncation error.
For these reasons, we select relatively large compression tolerance $\epsilon$ in order to achieve the reconstruction error of 
approximately $10^{-4}$. It is much larger than the truncation and the round-off errors.

To quantify the reconstruction accuracy of a vector field using a single scalar-valued metric, the maximum relative $L_\infty$ error norm is selected among the three velocity components,
\begin{equation}
e_{vec} = \max{(\frac{||\check{u}-u||_\infty}{||u||_\infty},\frac{||\check{v}-v||_\infty}{||v||_\infty},\frac{||\check{w}-w||_\infty}{||w||_\infty})}
\end{equation}
The compressed file size $\Sigma_{vec}$ is calculated as the sum of the compressed file sizes divided
by the original storage size of the three velocity components in double precision, and the compression ratio is equal to
$r_{vec} = 1 / \Sigma_{vec}$.

\begin{table}
\centering
\caption{Evaluation of the effect of derived representations of the velocity vector field.}
\begin{tabular}{ l l l l l }
\hline
                       & $\Sigma_{vec}$ & $r_{vec}$ & $\epsilon$           & $e_{vec}$ \\ \hline
Velocity -- Cartesian (original)  & $3.3$\%                 & $30.6$             & $10^{-4}$            & $9.5 \times 10^{-5}$ \\
Velocity -- polar      & $4.3$\%                 & $23.2$             & $3 \times 10^{-4}$   & $9.9 \times 10^{-5}$ \\
Vorticity -- Cartesian & $5.2$\%                 & $19.1$             & $3.5 \times 10^{-5}$ & $10.1 \times 10^{-5}$ \\
\end{tabular}
\label{table_vector}
\end{table}

Note that, in all methods, $e_{vec}$ is calculated on the reconstructed Cartesian velocity components $\check{u}$, $\check{v}$, $\check{w}$, 
but the tolerance $\epsilon$ is set on the transformed field in the `Velocity -- polar' and `Vorticity -- Cartesian' methods.
Therefore, in the two latter cases, $\epsilon$ is iteratively varied until $e_{vec}$ becomes close to $10^{-4}$.
The final values of $e_{vec}$ are shown in the last column of table~\ref{table_vector},
and the corresponding $\epsilon$ is in the second last column.
From the values of $\Sigma_{vec}$ and $r_{vec}$ in, respectively, the second and the third columns in table~\ref{table_vector} we conclude that 
the original `Velocity -- Cartesian' method provides the best compression for the desired $10^{-4}$ accuracy.

Such inefficiency of the `Velocity -- polar' and `Vorticity -- Cartesian' representations can be explained by 
the loss of uniformity in the spatial distribution of pointwise errors, and unequal errors of $\check{u}$, $\check{v}$ and $\check{w}$.
This implies that some point values are stored with higher precision than necessary.
In addition, the `Velocity -- polar' representation suffers from the discontinuity in $\vartheta$ and $\phi$ artificially introducing small scales in the field,
which require more wavelet coefficients to be stored.
To find a suitable vector field transform may be a promising direction for the future work.

\section{\label{sec:error_control}Reliability of the error control}

The wavelet transform offers control over the $L^2$ norm,
but we aim for the $L^\infty$ error control.
The poinwise error
in the physical space is a linear combination of the quantization errors of the wavelet coefficients
within the stencil. The latter are all smaller than $\varepsilon_F$ by construction. The weights are constants specific to the CFD9/7 wavelet.
The number of terms in the sum is bounded by the number of operations in the lifting steps times the number of spatial dimensions ($=3$) times the
number of levels of the transform ($=4$). From these considerations, one may derive a theoretical upper 
bound on the ratio between the $L^\infty$ error in physical space and the filtering threshold $\varepsilon_F$,
and thus obtain a theoretical estimate for the parameter $\eta$ in (\ref{eq:tolabs}).
However, we do not attempt such rigorous analysis. The value $\eta=1.75$ is an empirical constant.
Moreover, it does not exactly guarantee that $e_\infty = \varepsilon$.

\begin{figure}
    \begin{center}
        \includegraphics[scale=0.7]{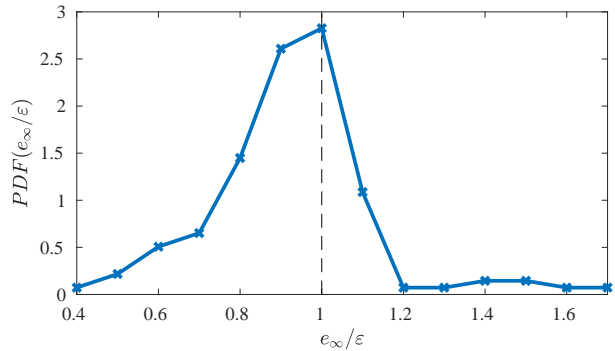}          
    \end{center}
    \caption{Histogram of the error-to-tolerance ratio.}
    \label{fig:pdf_errtol}
\end{figure}

In order to gain quantitative information about reliability of the error control in WaveRange across different application scenarios,
we compiled the $L^\infty$ error data $e_\infty$ from all examples (HIT, wake turbulence, typhoon, urban-scale and seismology simulations)
with different tolerance values $\varepsilon$ into one dataset. The error-to-tolerance ratio $e_\infty/\varepsilon$ was then calculated for each sample in the dataset. For the HIT and the wake turbulence, $e_\infty$ was taken separately for each velocity component. For the typhoon, urban-scale and seismology simulations,
the maximum $e_\infty$ over all fields was taken. For the urban-scale simulation, the united and the divided storage schemes were both included in the analysis.
This yielded a set of $186$ values of $e_\infty/\varepsilon$. We then discarded those samples that corresponded to $\varepsilon \le 10^{-14}$, for which
the error control failed because of the round-off errors. The remaining set contained $138$ samples.
We then calculated a histogram of $e_\infty/\varepsilon$. The result is shown in figure~\ref{fig:pdf_errtol}.

The peak of the histogram is at $e_\infty/\varepsilon = 1$. 
The expected value of $e_\infty/\varepsilon$ is estimated as $0.93$, and the standard deviation is equal to $0.20$.
By numerical integration of $PDF(e_\infty/\varepsilon)$ we found that, in $83\%$ of the cases included in the analysis, $e_\infty/\varepsilon$ was less or equal $1$. 
In $94\%$ cases it was less or equal $1.2$ and in no case it exceeded 1.7. 




\bibliographystyle{elsarticle-num} 
\bibliography{bib_bigdata_prf}

\begin{thebibliography}{10}
\expandafter\ifx\csname url\endcsname\relax
  \def\url#1{\texttt{#1}}\fi
\expandafter\ifx\csname urlprefix\endcsname\relax\def\urlprefix{URL }\fi
\expandafter\ifx\csname href\endcsname\relax
  \def\href#1#2{#2} \def\path#1{#1}\fi

\bibitem{Fornari_2016_pof}
W.~Fornari, A.~Formenti, F.~Picano, L.~Brandt, Phys.\ Fluids 28~(3) (2016)
  033301.
\newblock \href {http://dx.doi.org/10.1063/1.4942518}
  {\path{doi:10.1063/1.4942518}}.

\bibitem{Bakke_2016_aa}
H.~Bakke, L.~Frogner, B.~V. Gudiksen, Astron.\ Astrophys. 620 (2018) L5.
\newblock \href {http://dx.doi.org/10.1051/0004-6361/201834129}
  {\path{doi:10.1051/0004-6361/201834129}}.

\bibitem{Baudron_2007_nse}
A.-M. Baudron, J.-J. Lautard, Nucl.\ Sci.\ Eng. 155~(2) (2007) 250--263.
\newblock \href {http://dx.doi.org/10.13182/NSE07-A2660}
  {\path{doi:10.13182/NSE07-A2660}}.

\bibitem{Kageyama_2004_ggg}
A.~Kageyama, T.~Sato, Geochem.\ Geophys. 5~(9).
\newblock \href {http://dx.doi.org/10.1029/2004GC000734}
  {\path{doi:10.1029/2004GC000734}}.

\bibitem{Ronchi_1996_jcp}
C.~Ronchi, R.~Iacono, P.~S. Paolucci, J.\ Comput.\ Phys. 124~(1) (1996)
  93--114.
\newblock \href {http://dx.doi.org/10.1006/jcph.1996.0047}
  {\path{doi:10.1006/jcph.1996.0047}}.

\bibitem{Nakano_2017_gmd}
M.~Nakano, A.~Wada, M.~Sawada, H.~Yoshimura, R.~Onishi, S.~Kawahara, W.~Sasaki,
  T.~Nasuno, M.~Yamaguchi, T.~Iriguchi, M.~Sugi, Y.~Takeuchi, Geosci.\ Model
  Dev. 10~(3) (2017) 1363--1381.
\newblock \href {http://dx.doi.org/10.5194/gmd-10-1363-2017}
  {\path{doi:10.5194/gmd-10-1363-2017}}.

\bibitem{Kageyama_2004_proc}
A.~Kageyama, M.~Kameyama, S.~Fujihara, M.~Yoshida, M.~Hyodo, Y.~Tsuda, A 15.2
  {TF}lops simulation of geodynamo on the {E}arth {S}imulator, in: SC '04:
  Proceedings of the 2004 ACM/IEEE Conference on Supercomputing, 2004, pp.
  35--35.
\newblock \href {http://dx.doi.org/10.1109/SC.2004.1}
  {\path{doi:10.1109/SC.2004.1}}.

\bibitem{Komatitsch_2013_book}
D.~Komatitsch, S.~Tsuboi, J.~Tromp, The spectral-element method in seismology,
  American Geophysical Union (AGU), 2013, pp. 205--227.
\newblock \href {http://dx.doi.org/10.1029/157GM13}
  {\path{doi:10.1029/157GM13}}.

\bibitem{Korpilo_2016_cpc}
T.~Korpilo, A.~D. Gurchenko, E.~Z. Gusakov, J.~A. Heikkinen, S.~J. Janhunen,
  T.~P. Kiviniemi, S.~Leerink, P.~Niskala, A.~A. Perevalov, Comput.\ Phys.\
  Commun. 203 (2016) 128--137.
\newblock \href {http://dx.doi.org/10.1016/j.cpc.2016.02.021}
  {\path{doi:10.1016/j.cpc.2016.02.021}}.

\bibitem{Hilbert_2016}
M.~Hilbert, Dev.\ Pol.\ Rev. 34 (2016) 135--174.

\bibitem{Taubman_2002_book}
D.~Taubman, M.~Marcellin, {JPEG2000} image compression fundamentals, standards
  and practice, 1st Edition, Vol. 642 of The Springer International Series in
  Engineering and Computer Science, Springer US, 2002.

\bibitem{Schmalzl_2003_cg}
J.~Schmalzl, Comput.\ Geosci. 29~(8) (2003) 1021--1031.
\newblock \href {http://dx.doi.org/10.1016/S0098-3004(03)00098-0}
  {\path{doi:10.1016/S0098-3004(03)00098-0}}.

\bibitem{Woodring_2011_ieee}
J.~Woodring, S.~Mniszewski, C.~Brislawn, D.~De{M}arle, J.~Ahrens, Revisiting
  wavelet compression for large-scale climate data using {JPEG} 2000 and
  ensuring data precision, in: 2011 IEEE Symposium on Large Data Analysis and
  Visualization, 2011, pp. 31--38.
\newblock \href {http://dx.doi.org/10.1109/LDAV.2011.6092314}
  {\path{doi:10.1109/LDAV.2011.6092314}}.

\bibitem{Peyrot_2019_jcg}
J.-L. Peyrot, L.~Duval, F.~Payan, L.~Bouard, L.~Chizat, S.~Schneider,
  M.~Antonini, Comput.\ Geosci. 23~(4) (2019) 723--743.
\newblock \href {http://dx.doi.org/10.1007/s10596-019-9816-2}
  {\path{doi:10.1007/s10596-019-9816-2}}.

\bibitem{Sakai_2011_conf}
R.~Sakai, D.~Sasaki, K.~Nakahashi, Large-scale {CFD} data compression for
  building-cube method using wavelet transform, in: A.~Kuzmin (Ed.),
  Computational Fluid Dynamics 2010, Springer Berlin Heidelberg, Berlin,
  Heidelberg, 2011, pp. 465--470.

\bibitem{Sakai_2013_candf}
R.~Sakai, D.~Sasaki, K.~Nakahashi, Comput.\ Fluids 80 (2013) 116--127.

\bibitem{Sakai_2013_ijnmf}
R.~Sakai, D.~Sasaki, S.~Obayashi, K.~Nakahashi, Int.\ J.\ Numer.\ Meth.\ Fl.
  73~(5) (2013) 462--476.
\newblock \href {http://dx.doi.org/10.1002/fld.3808}
  {\path{doi:10.1002/fld.3808}}.

\bibitem{Hatfield_2018_mwr}
S.~Hatfield, A.~Subramanian, T.~Palmer, P.~D\"uben, Mon.\ Weather Rev. 146~(1)
  (2018) 49--62.

\bibitem{Baker_2016_gmd}
A.~H. Baker, D.~M. Hammerling, S.~A. Mickelson, H.~Xu, M.~B. Stolpe, P.~Naveau,
  B.~Sanderson, I.~Ebert-Uphoff, S.~Samarasinghe, F.~De~Simone, F.~Carbone,
  C.~N. Gencarelli, J.~M. Dennis, J.~E. Kay, P.~Lindstrom, Geosci.\ Model Dev.
  9~(12) (2016) 4381--4403.
\newblock \href {http://dx.doi.org/10.5194/gmd-9-4381-2016}
  {\path{doi:10.5194/gmd-9-4381-2016}}.

\bibitem{Laney_2013_ieee}
D.~Laney, S.~Langer, C.~Weber, P.~Lindstrom, A.~Wegener, Assessing the effects
  of data compression in simulations using physically motivated metrics, in: SC
  '13: Proceedings of the International Conference on High Performance
  Computing, Networking, Storage and Analysis, 2013, pp. 1--12.
\newblock \href {http://dx.doi.org/10.1145/2503210.2503283}
  {\path{doi:10.1145/2503210.2503283}}.

\bibitem{Farge_1992_arfm}
M.~Farge, Ann.\ Rev.\ Fluid Mech. 24~(1) (1992) 395--458.

\bibitem{Schneider_2010_arfm}
K.~Schneider, O.~V. Vasilyev, Ann.\ Rev.\ Fluid Mech. 42~(1) (2010) 473--503.

\bibitem{Bradley_1993_proc}
J.~N. Bradley, C.~M. Brislawn, Wavelet transform-vector quantization
  compression of supercomputer ocean models, in: [Proceedings] DCC '93: Data
  Compression Conference, 1993, pp. 224--233.
\newblock \href {http://dx.doi.org/10.1109/DCC.1993.253127}
  {\path{doi:10.1109/DCC.1993.253127}}.

\bibitem{Wilson_2002_proc}
J.~P. Wilson, Wavelet-based lossy compression of barotropic turbulence
  simulation data, in: Proceedings DCC 2002. Data Compression Conference, 2002,
  pp. 479--.
\newblock \href {http://dx.doi.org/10.1109/DCC.2002.1000022}
  {\path{doi:10.1109/DCC.2002.1000022}}.

\bibitem{Kang_2003_ksme}
H.~Kang, D.~Lee, D.~Lee, KSME Int.\ J. 17~(11) (2003) 1784--1792.
\newblock \href {http://dx.doi.org/10.1007/BF02983609}
  {\path{doi:10.1007/BF02983609}}.

\bibitem{Li_2015_ieee}
S.~Li, K.~Gruchalla, K.~Potter, J.~Clyne, H.~Childs, Evaluating the efficacy of
  wavelet configurations on turbulent-flow data, in: 2015 IEEE 5th Symposium on
  Large Data Analysis and Visualization (LDAV), 2015, pp. 81--89.
\newblock \href {http://dx.doi.org/10.1109/LDAV.2015.7348075}
  {\path{doi:10.1109/LDAV.2015.7348075}}.

\bibitem{Berkooz_1993_arfm}
G.~Berkooz, P.~Holmes, J.~L. Lumley, Ann.\ Rev.\ Fluid Mech. 25~(1) (1993)
  539--575.

\bibitem{Balajewicz_2013_jfm}
M.~J. Balajewicz, E.~H. Dowell, B.~R. Noack, J.\ Fluid Mech. 729 (2013)
  285--308.

\bibitem{Schmid_2010_jfm}
P.~J. Schmid, J.\ Fluid Mech. 656 (2010) 5--28.

\bibitem{Lorente_2010_ast}
L.~S. Lorente, J.~M. Vega, A.~Velazquez, Aerosp.\ Sci.\ Technol. 14~(3) (2010)
  168--177.

\bibitem{Bi_2014_ieee}
C.~Bi, K.~Ono, L.~Yang, Parallel {POD} compression of time-varying big datasets
  using m-swap on the {K} computer, in: 2014 IEEE International Congress on Big
  Data, 2014, pp. 438--445.

\bibitem{Schlegel_2009_proc}
M.~Schlegel, B.~R. Noack, P.~Comte, D.~Kolomenskiy, K.~Schneider, F.~M., D.~M.
  Luchtenburg, J.~E. Scouten, G.~Tadmor, Reduced-order modelling of turbulent
  jets for noise control, in: D.~Juv\'e, M.~Manhart, C.~D. Munz (Eds.),
  Numerical Simulation of Turbulent Flows and Noise Generation, Vol. 104 of
  Notes on Numerical Fluid Mechanics and Multidisciplinary Design, Springer
  Berlin Heidelberg, Berlin, Heidelberg, 2009, pp. 3--27.

\bibitem{Rosten_2004_gp}
T.~R{\o}sten, T.~A. Ramstad, L.~Amundsen, Geophys.\ Prospect. 52~(5) (2004)
  359--378.
\newblock \href {http://dx.doi.org/10.1111/j.1365-2478.2004.00422.x}
  {\path{doi:10.1111/j.1365-2478.2004.00422.x}}.

\bibitem{Duval_2000_proc}
L.~Duval, T.~R{\o}sten, Filter bank decomposition of seismic data with
  application to compression and denoising, in: SEG Annual International
  Meeting, Soc. Expl. Geophysicists, 2000, pp. 2055--2058.
\newblock \href {http://dx.doi.org/10.1190/1.1815847}
  {\path{doi:10.1190/1.1815847}}.

\bibitem{Najmabadi_2017_comput}
S.~M. Najmabadi, P.~Offenh\"{a}user, M.~Hamann, G.~Jajnabalkya, F.~Hempert,
  C.~W. Glass, S.~Simon, Computation 5~(2).
\newblock \href {http://dx.doi.org/10.3390/computation5020024}
  {\path{doi:10.3390/computation5020024}}.

\bibitem{Lakshminarasimhan_2011_europar}
S.~Lakshminarasimhan, N.~Shah, S.~Ethier, S.~Klasky, R.~Latham, R.~Ross, N.~F.
  Samatova, Compressing the incompressible with {ISABELA}: In-situ reduction of
  spatio-temporal data, in: Euro-Par 2011 Parallel Processing, Lecture Notes in
  Computer Science, Springer Berlin Heidelberg, 2011, pp. 366--379.

\bibitem{Liang_2018_ieee}
X.~Liang, S.~Di, D.~Tao, S.~Li, S.~Li, H.~Guo, Z.~Chen, F.~Cappello, 2018 IEEE
  International Conference on Big Data (Big Data) (2018) 438--447.

\bibitem{Li_2018_cgf}
S.~Li, N.~Marsaglia, C.~Garth, J.~Woodring, J.~Clyne, H.~Childs, Comput.\
  Graph.\ Forum 37~(6) (2018) 422--447.
\newblock \href {http://dx.doi.org/10.1111/cgf.13336}
  {\path{doi:10.1111/cgf.13336}}.

\bibitem{Lundquist_2010_mwr}
K.~A. Lundquist, F.~K. Chow, J.~K. Lundquist, Mon.\ Weather Rev. 138~(3) (2010)
  796--817.

\bibitem{Matsuda_2018_jweia}
K.~Matsuda, R.~Onishi, K.~Takahashi, J.\ Wind Eng.\ Ind.\ Aerod. 173 (2018)
  53--66.

\bibitem{Daubechies_1998_jfaa}
I.~Daubechies, W.~Sweldens, J.\ Fourier Anal.\ Appl. 4~(3) (1998) 247--269.

\bibitem{Getreuer_website}
P.~Getreuer, Wavelet {CDF} 9/7 implementation, retrieved from
  \url{http://www.getreuer.info/home/waveletcdf97/}, \today (2013).

\bibitem{Shannon_1948_bell}
C.~E. Shannon, ‎Bell Syst.\ Tech.\ J. 27 (1948) 379--423, 623--656.

\bibitem{Martin_1979_conf}
G.~N.~N. Martin, Range encoding: an algorithm for removing redundancy from a
  digitized message, in: Video \& Data Recording Conference, Southampton, UK,
  1979.

\bibitem{Schindler_1999_website}
M.~Schindler, Range encoder homepage, retrieved from
  \url{http://www.compressconsult.com/rangecoder/}, \today (1999).

\bibitem{Engels_2016_sisc}
T.~Engels, D.~Kolomenskiy, K.~Schneider, J.~Sesterhenn, SIAM J.\ Sci.\ Comput.
  38 (2016) S3--S24.

\bibitem{Takahashi_2013_jop}
K.~Takahashi, R.~Onishi, Y.~Baba, S.~Kida, K.~Matsuda, K.~Goto, H.~Fuchigami,
  J.\ Phys.\ Conf.\ Ser. 454~(1) (2013) 012072.

\bibitem{Folk_1999_ps}
M.~Folk, A.~Cheng, K.~Yates, {HDF5}: {A} file format and {I/O} library for high
  performance computing applications, in: Proceedings of Supercomputing,
  Vol.~99, Portland, Oregon, 1999, pp. 5--33.

\bibitem{Onishi_2013_jcp}
R.~Onishi, K.~Takahashi, J.~C. Vassilicos, J.\ Comput.\ Phys. 242 (2013)
  809--827.

\bibitem{Onishi_2011_jcp}
R.~Onishi, Y.~Baba, K.~Takahashi, J.\ Comput.\ Phys. 230~(10) (2011)
  4088--4099.

\bibitem{Onishi_2012_jas}
R.~Onishi, K.~Takahashi, J.\ Atmos.\ Sci. 69~(5) (2012) 1474--1497.
\newblock \href {http://dx.doi.org/10.1175/JAS-D-11-0166.1}
  {\path{doi:10.1175/JAS-D-11-0166.1}}.

\bibitem{Sekiguchi_2008_jqsrt}
M.~Sekiguchi, T.~Nakajima, J.\ Quant.\ Spectrosc.\ Ra. 109~(17) (2008)
  2779--2793.
\newblock \href {http://dx.doi.org/10.1016/j.jqsrt.2008.07.013}
  {\path{doi:10.1016/j.jqsrt.2008.07.013}}.

\bibitem{Deutsch_1996_techrep}
P.~Deutsch, {DEFLATE} compressed data format specification version 1.3, Tech.
  rep. (1996).

\bibitem{Yeh_2002_estc}
P.-S. Yeh, W.~Xia-Serafino, L.~Miles, B.~Kobler, D.~Menasce, Implementation of
  {CCSDS} lossless data compression in {HDF}, in: Earth Science Technology
  Conference -- 2002, Pasadena, California, 2002, p. A3P2.

\bibitem{Pavlov_2019_techrep}
I.~Pavlov, 7z format, retrieved from \url{http://www.7-zip.org/7z.html}, \today
  (2019).

\bibitem{Tao_2017_ieee}
D.~Tao, S.~Di, Z.~Chen, F.~Cappello, Significantly improving lossy compression
  for scientific data sets based on multidimensional prediction and
  error-controlled quantization, in: Proceedings of the 31st {IEEE}
  International Parallel and Distributed Processing Symposium (IPDPS), Orlando,
  Florida, 2017, pp. 1129--1139.

\bibitem{Diffenderfer_2019_sisc}
J.~Diffenderfer, A.~Fox, J.~Hittinger, G.~Sanders, P.~Lindstrom, SIAM J.\ Sci.\
  Comput. 41~(3) (2019) A1867--A1898.
\newblock \href {http://dx.doi.org/10.1137/18M1168832}
  {\path{doi:10.1137/18M1168832}}.

\bibitem{Ravi_2016_srep}
S.~Ravi, D.~Kolomenskiy, T.~Engels, K.~Schneider, C.~Wang, J.~Sesterhenn,
  H.~Liu, Sci.\ Rep. 6 (2016) 35043.

\end{thebibliography}





\clearpage




\section*{Current code version}
\label{}

\begin{table}[!h]
\begin{tabular}{|l|p{6.5cm}|p{6.5cm}|}
\hline
\textbf{Nr.} & \textbf{Code metadata description} & \textbf{Code metadata value} \\
\hline
C1 & Current code version & v3.15.3 \\
\hline
C2 & Permanent link to code/repository used of this code version & {\tiny \url{https://github.com/pseudospectators/WaveRange/releases/tag/v3.15.3}} \\
\hline
C3 & Legal Code License   & GPL-3.0 \\
\hline
C4 & Code versioning system used & git \\
\hline
C5 & Software code languages, tools, and services used & c, c++ \\
\hline
C6 & Compilation requirements, operating environments \& dependencies & Linux \\
\hline
C7 & If available Link to developer documentation/manual & {\tiny \url{https://github.com/pseudospectators/WaveRange/blob/master/README.md}} \\
\hline
C8 & Support email for questions & \url{dkolom@gmail.com} \\
\hline
\end{tabular}
\caption{Code metadata}
\label{} 
\end{table}

\end{document}